\documentclass[12pt,journal,draftclsnofoot,oneside,onecolumn]{IEEEtran}
\usepackage{subfigure}
\usepackage[cmex10]{amsmath}
\usepackage{amssymb}
\usepackage{epsfig}
\usepackage{cite}
\usepackage{graphicx}
\usepackage{color}
\usepackage{bm}
\usepackage{booktabs}
\usepackage{gensymb}
\usepackage{mathrsfs}
\usepackage{xfrac}
\usepackage{algorithm}
\usepackage{algorithmic}
\usepackage{multirow}
\usepackage{float}
\newfloat{figtab}{htb}{fgtb}
\makeatletter
 \newcommand\figcaption{\def\@captype{figure}\caption}
  \newcommand\tabcaption{\def\@captype{table}\caption}

\newlength{\figwidth}
\newcommand{\tabincell}[2]{\begin{tabular}{@{}#1@{}}#2\end{tabular}}

\newtheorem{remark}{\it Remark}
\newtheorem{proposition}{\it Proposition}


\begin{document}

\title{ Simultaneous Wireless Information and Power Transfer in Near-Field Communications} 
\author{Zheng Zhang,~\IEEEmembership{Student Member,~IEEE},  Yuanwei Liu,~\IEEEmembership{Senior Member,~IEEE},  Zhaolin Wang,~\IEEEmembership{Graduate Student Member,~IEEE}, \\ Xidong Mu,~\IEEEmembership{Member,~IEEE}, and Jian Chen,~\IEEEmembership{Member,~IEEE},\vspace{-6mm}
\thanks{Zheng Zhang and Jian Chen are with the School of Telecommunications Engineering, Xidian University, Xi'an 710071, China (e-mail: zzhang\_688@stu.xidian.edu.cn; jianchen@mail.xidian.edu.cn). Yuanwei Liu, Zhaolin Wang and Xidong Mu are with the School of Electronic Engineering and Computer Science,
Queen Mary University of London, London E1 4NS, U.K. (e-mail: yuanwei.liu@qmul.ac.uk; zhaolin.wang@qmul.ac.uk; xidong.mu@qmul.ac.uk;).}
}
\maketitle

\begin{abstract}
  A near-field simultaneous wireless information and power transfer (SWIPT) network is investigated, where the hybrid beamforming architecture is employed at the base station (BS) for information transmission while charging energy harvesting users. A transmit power minimization problem is formulated by jointly optimizing of the analog beamformer, the baseband digital information/energy beamformers, and the number of dedicated energy beams. To tackle the uncertain number of dedicated energy beams, a semidefinite relaxation based rank-one solution construction method is proposed to obtain the optimal baseband digital beamformers under the fixed analog precoder. Based on the structure of the optimal baseband digital beamformers, it is proved that no dedicated energy beam is required in near-field SWIPT. To further exploit this insight, a penalty-based two-layer (PTL) algorithm is proposed to optimize the analog beamformer and baseband digital information beamformers. By employing the block coordinate descent method, the optimal analog beamformer and baseband digital information beamformers are obtained in the closed-form expressions. Moreover, to reduce the high computational complexity caused by the large number of antennas, a low-complexity two-stage algorithm is proposed. Numerical results illustrate that: 1) the proposed PTL algorithm can achieve the near-optimal performance; and 2) in contract to the far-field SWIPT, single near-field beamformer can focus the energy on multiple locations.
\end{abstract}

\IEEEpeerreviewmaketitle

\section{Introduction}
To cope with the explosive growth of the number of the Internet-of-Things (IoT) devices, numerous promising technologies, e.g., multiple-input-multiple-output (MIMO), millimeter wave (mmWave), and reconfigurable intelligent surface (RIS), have been proposed to enable the ubiquitous IoT connectivity \cite{F.Boccardi_5G_magazine,6G_samsung,Y.Liu_RIS}. However, the huge number and wide distribution of IoT devices make it incredibly challenging and costly to provide a seamless and reliable power supply for IoT devices. To deal with this issue, the dual use of radio frequency (RF) signals for simultaneous wireless information and power transfer (SWIPT) has gained significant attention \cite{I.Krikidis_SWIPT}. By integrating the RF energy harvesting ability into the IoT devices, it is expected to provide a permanent and highly controllable energy supply for the IoT devices \cite{L.Liu_SWIPT,S.Bi_WPT_mag}.

On the other hand, deploying an extremely large-scale antenna array (ELAA) at the base station (BS) has become a common trend in future wireless networks, which is envisioned as a key enabler for achieving ultra-high spectral efficiency, energy efficiency, and reliability. Nevertheless, increasing antenna aperture inevitably expands the near-field region of the wireless networks, where the electromagnetic (EM) propagation characteristics are fundamentally changed \cite{Y.Liu_NF_mag}. Specifically, the angular field becomes correlated with the distance to the receiver, and the accurate spherical-wave channel model should be adopted \cite{J.Xu_NF,C.Huang_NF_channel,R.Ji_DoF}. By exploiting the extra distance information contained in the spherical-wave channel, the transmit array radiation patterns can achieve the energy focus on a targeted point in the near field, which is referred to as \textit{beamfocusing} \cite{H.Zhang_NF_mag,H.Zhang_NF_mag2}. Thus, the unique near-field effect provides more flexibility for spatial beamforming design, which has become a non-negligible factor in future wireless networks \cite{M.Cui_NF_mag,C.Huang_NF1}.

\subsection{Prior Works}

\subsubsection{Studies on SWIPT}
Extensive research has been devoted to SWIPT \cite{L.R.Varshney_SWIPT,R.Zhang_SWIPT,J.Xu_energy_beam1,Q.Wu_SWIPT_wcl,Y.Liu_SWIPT_NOMA}. For example, the authors of \cite{L.R.Varshney_SWIPT} introduced the concept of SWIPT, in which a capacity-energy function was defined to evaluate the rate-energy (R-E) trade-off for single-input single-output (SISO) networks. To facilitate the hardware implementation of SWIPT, the authors of \cite{R.Zhang_SWIPT} investigated a MIMO broadcast channel SWIPT network, where two practical receiving architectures, i.e., time-switching (TS) and power-splitting (PS), were designed. Considering the spatial beamforming design at the transmitters, a general multi-user SWIPT scenario was considered in \cite{J.Xu_energy_beam1}, in which the authors rigorously proved that the information beams are sufficient to provide optimal performance for the SWIPT network. To introduce more degrees-of-freedom (DoFs), the authors of \cite{Q.Wu_SWIPT_wcl} proposed to apply the RIS to the SWIPT network, where joint active and passive beamforming optimization schemes were devised to enhance energy efficiency, where the non-necessity of the dedicated energy beams in the RIS-enabled SWIPT network were proved. Moreover, it was claimed that the non-orthogonal multiple access (NOMA) could be integrated into the SWIPT networks to provide a win-win solution in terms of spectrum efficiency and energy efficiency \cite{Y.Liu_SWIPT_NOMA}.

\subsubsection{Studies on near-filed communications}
With the significant rise of the antenna array scale, near-field communications have drawn a great deal of attention recently \cite{H.Zhang_NF,LDMA,NF_analysis,H.Zhang_channel_estimation,C.You_NF_training,C.Wu_NF_training}.
Specifically, the authors of \cite{H.Zhang_NF} first revealed the possibility of the beamfocusing in near-field communications, where the distance information contained in spherical wave channels was exploited to focus the signal on a specific location. As an advance, the authors of \cite{LDMA} proposed a novel location division multiple access (LDMA) scheme. The asymptotic orthogonality of the spherical wave array response vectors in the distance dimension was rigorously proved, which demonstrated the spectral potential of the near-field communications. Following the above works, the authors of \cite{NF_analysis} designed a general framework for near-field analysis, where the impact of the discrete array aperture on the received power was unveiled. However, the extremely large-scale antennas of near-field communications imposed enormous pilot overheads on the BS, which motivates the authors of \cite{H.Zhang_channel_estimation} to design the low-complexity channel estimation method by using the sparsifying dictionary of near-field channels. Furthermore, the authors of \cite{C.You_NF_training,C.Wu_NF_training} proposed to employ the beam training based approaches to jointly optimize the network performance without channel state information (CSI), where the hybrid array architecture is adopted to further reduce the RF chain costs.

\subsubsection{Studies on near-filed WPT}
To counter the low wireless power transfer (WPT) efficiency over long distances, there have been a few works focusing on integrating the ELAA into the WPT-empowered networks for energy efficiency enhancement \cite{H.Zhang_WPT,E.Demarchou_energy_focusing,H.Zhang_NF_ISWPT}. Particularly, the authors of \cite{H.Zhang_WPT} proposed a dynamic metasurface antenna based near-field WPT framework, where an efficient Riemannian conjugate gradient approach was developed to enable the energy beamfocusing. To further explore the characterization of near-field energy harvesting, the work of \cite{E.Demarchou_energy_focusing} designed a general analysis framework, where the closed-form expressions of the harvested power were derived for the fixed and random locations, respectively. As a further advance, the authors of \cite{H.Zhang_NF_ISWPT} proposed a novel concept of integrating sensing and wireless power transfer for near-field communication, where the fundamental trade-off between the WPT and radar sensing was characterized via optimizing the transmit beamforming. 


\subsection{Motivations and Contributions}
Although a few of works have been devoted to the near-field WPT-enabled networks \cite{H.Zhang_WPT,E.Demarchou_energy_focusing,H.Zhang_NF_ISWPT}, the investigation of near-field SWIPT is still in its infancy, which have many open problems and challenges to be addressed.
\begin{itemize}
  \item On the one hand, the fundamental problem in near-filed SWIPT is that whether the dedicated energy beams are required. Compared to the conventional far-field SWIPT, the spherical-wave-based near-field propagation enables the unique beamfocusing targeting specific locations. The intuition is that since the near-field beamfocusing provides superior capability of desired signal enhancement and interference mitigation, the dedicated energy beams are indeed necessary to achieve the optimal SWIPT performance rather than merely relying on the information beams \cite{H.Zhang_NF_mag2}. 

  \item On the other hand, for facilitating near-field SWIPT, it inevitably requires an extremely large scale of antennas to be employed. To address the potentially high signal processing overhead at the transmitter, tailored low-complexity beamfocusing SWIPT scheme has to be developed for practical implementation. 

\end{itemize}

Against the above background, in this paper, we investigate the near-field SWIPT beamfocusing design, where all the information decoding users (IDs) and the energy harvesting users (EHs) are located in the spherical-wave-based near-field region. In particular, we aim to address the fundamental issue of whether the dedicated energy beams are essential for near-field SWIPT and develop low-complexity beamfocusing method for efficiently operating the near-field SWIPT. Our main contributions are summarized as follows.
\begin{itemize}
  \item We investigate a near-field SWIPT network, in which the BS employs the large-scale hybrid beamforming architecture to send information-embedded and energy-charging signals to the near-field IDs and EHs. We formulate an optimization problem to minimize the total transmit power at the BS by jointly optimizing the analog beamformer, the baseband digital information and energy beamformers, and the number of dedicated energy beams, subject to the quality-of-service (QoS) requirements of the IDs and the RF receive power constraints at the EHs.
  \item To obtain insights into the impact of the number of the dedicated energy beams on near-field SWIPT networks, we study the baseband digital beamformers design subporblem under any given analog beamformer, which is solved by exploiting the semidefinite relaxation (SDR). We propose a generic approach for constructing a optimal rank-one solution of the baseband digital beamformers that guarantees the tightness of SDR. Based on the structure of this optimal solution, we rigorously prove that the optimal dedicated energy beamformer matrix has a rank of zero. It indicates that the dedicated energy beams are not required for near-field SWIPT.
  \item Based on above insight, we focus on the joint optimization of the baseband digital information beamformers and the analog beamformer. To address the coupling in the resulting optimization problem, we propose a penalty-based two-layer (PTL) algorithm. By employing the block coordinate descent (BCD) framework, the analog beamformer and baseband digital beamformers are derived in the closed-form expressions, respectively. Furthermore, a low-complexity two-stage hybrid beamforming algorithm is developed, whose computational complexity only relies on the small-scale number of employed RF chains, thus being attractive for the practical implementation.
  \item Simulation results validate the correctness of the obtained conclusion. It is demonstrated that: 1) the proposed PTL algorithm achieves the near-optimal SWIPT performance; 2) the proposed low-complexity two-stage algorithm is only slightly inferior to the PTL algorithm but enjoys a significant low complexity; and 3) single near-field beamformer is able to achieve the energy focusing on multiple locations, thus benefiting both IDs and EHs.
\end{itemize}

\subsection{Organization and Notations}
The remainder of this paper is as follows. The system model and problem formulations are presented in Section \ref{Section_2}. Section \ref{Section_3} rigorously proves the non-necessity of the dedicated energy beams in near-field SWIPT networks. In Section \ref{Section_4}, a PTL algorithm and a low-complexity two-stage algorithm are conceived for the joint optimization of analog beamformer and baseband digital beamformers. The numerical results are presented in Section \ref{Section_5}. Finally, the conclusion are drawn in Section \ref{Section_6}.

\textit{Notations:} The boldface capital $\mathbf{X}$, the boldface lower-case letter $\mathbf{x}$, and the lower-case letter $x$ denote the matrix, veector, and scalar, respectively. $\mathbf{X}\in\mathbb{C}^{N\times M}$ denotes a matrix $\mathbf{X}$ with $N$ rows and $M$ columns, while $\mathbf{X}^{T}$ and $\mathbf{X}^{H}$ denote the transpose and Hermitian conjugate operations of $\mathbf{X}$, respectively. Also, $\text{rank}(\mathbf{X})$, $\text{Tr}(\mathbf{X})$, $\|\mathbf{X}\|$, $\|\mathbf{X}\|_{\text{F}}$, $\text{vec}(\mathbf{X})$, and $\mathbf{X}^{-1}$ represent the rank value, trace value, spectral norm, Frobenius norm, vectorized vector, and the inverse matrix of the matrix $\mathbf{X}$. Furthermore, the representation of $\mathbf{X}\succeq\mathbf{0}$ guarantees that $\mathbf{X}$ is positive semidefinite, and $\mathbf{X}^{[i,j]}$ denotes the $i$-th row and the $j$-column element of $\mathbf{X}$. The constant identity matrix and zero matrix are respectively denoted by $\mathbf{I}$ and $\mathbf{0}$. For any given vector $\mathbf{x}$, $\mathbf{x}\sim \mathcal{CN}(0,\mathbf{X})$ denotes that  $\mathbf{x}$ follows circularly symmetric complex Gaussian (CSCG) distribution with zero mean and covariance matrix $\mathbf{X}$. Similarly, $\text{diag}(\mathbf{x})$, $\|\mathbf{x}\|$, and $|\mathbf{x}|$ denote the diagonal matrix, Euclidean norm, and the absolute-value vector of $\mathbf{x}$. The notations of $\mathbb{E}($\textperiodcentered$)$ and $\Re(\cdot)$ are used to represent the statistical expectation and real component of the corresponding variable. $\odot$ is used to denote the Hadamard product.

\section{System Model and Problem Formulation}\label{Section_2}

\begin{figure}[h]
  \centering
  \includegraphics[scale = 0.5]{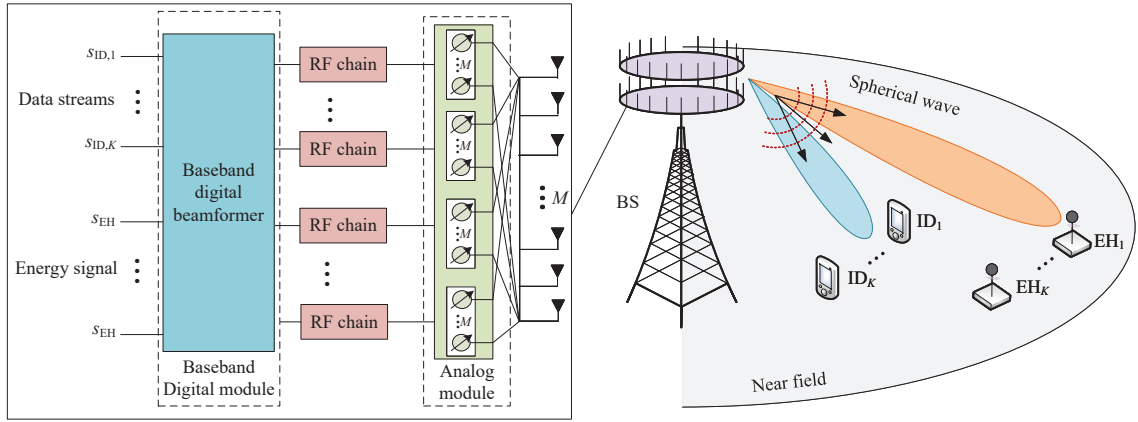}
  \caption{The near-field SWIPT network.}
  \label{Fig.1}
\end{figure}

We consider a near-field SWIPT system, as shown in the right half of Fig. \ref{Fig.1}. The BS equipped with an $M$-antenna uniform linear array (ULA) broadcasts wireless signal to $K$ single-antenna IDs and $L$ single-antenna EHs. Let us denote the sets of IDs and EHs as $\mathcal{K}_{\text{ID}}\triangleq\{1,\cdots,K\}$ and $\mathcal{K}_{\text{EH}}\triangleq\{1,\cdots,L\}$. All IDs and EHs are assumed to be located in the near-field region. In other words, the distances between the BS and IDs/EHs are shorter than the Rayleigh distance $d_{\text{R}}=\frac{2D_{1}^{2}}{\lambda}$, where $\lambda$ and $D$ denote the signal wavelength and the antenna aperture of the BS, respectively. However, near-field communications usually happen in the high-frequency and ELAA networks, which impose significant hardware challenges on the RF chain deployment. Specifically, the low noise amplifier and power amplifier inside the RF chain should be squeezed into the back of each antenna, whereas the antennas are positioned extremely close to each other for high main-lobe gain. Thus, it becomes difficult to allocate individual RF chain to each antenna. To tackle this issue, we adopt the hybrid beamforming structure in this paper, as shown in the left half of Fig. \ref{Fig.1}. In particular, the phase-shift based analog beamformer is deployed between the $M_{\text{RF}}$ ($K+L\leq M_{\text{RF}}\ll M$) RF chains and $M$ transmit antennas, where the output of each RF chain is connected to all the transmit antennas.
\subsection{Near-Field Channel Model}

\begin{figure}[t]
  \centering
  \includegraphics[scale = 0.7]{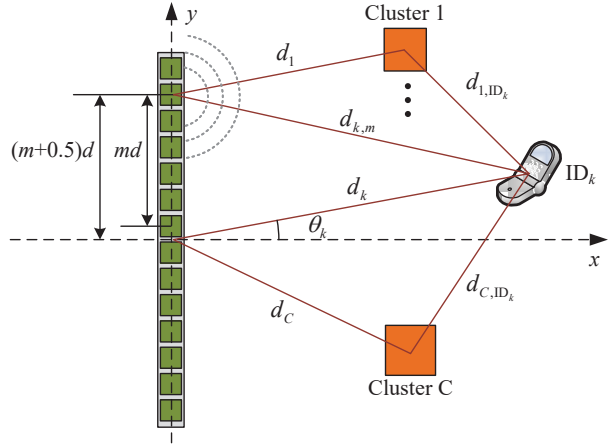}
  \caption{The near-field spherical wave channel model.}
  \label{Fig_NF_channel}
\end{figure}

As shown in Fig. \ref{Fig_NF_channel}, we consider a multi-path near-field channel model for the IDs. Without loss of generality, we define the origin of the coordinate system as the midpoint of the ULA at the BS and set the y-axis along the ULA. Therefore, the coordinate of the $m$-th element of the ULA is given by $\mathbf{s}_m = (0, \tilde{m} d)$, $d$ denotes the antenna spacing, and $\tilde{m}=\frac{M}{2}-m+\frac{d}{2}$. Let $d_{k}$ and $\theta_{k}$ denote the distance and angle of the $\text{ID}_{k}$ with respect to the original of the coordinate system. The coordinate of the $\text{ID}_{k}$ is thus given by $\mathbf{r}_k = (d_{k} \cos \theta_{k}, d_{k} \sin \theta_k)$. Therefore, the distance between the $m$-th element of the ULA and the $\text{ID}_{k}$ is given by
\begin{equation}\label{1}
d_{k,m} = \|\mathbf{r}_k - \mathbf{s}_m\| = \sqrt{d_{k}^{2}+(\tilde{m}d)^{2}-2\tilde{m}d d_{k}\sin\theta_{k}}.
\end{equation}
According to \cite{LDMA}, the LoS near-field channel between the BS and the $\text{ID}_{k}$ can be expressed as follows:
\begin{equation}\label{2}
\mathbf{h}_{\text{ID},k}^{\text{LoS}}=g_{k}\mathbf{a}(d_{k},\theta_{k}),
\end{equation}
where $a(d,\theta)$ denotes the near-field array response vector and is given by
\begin{equation}\label{3}
\mathbf{a}(d, \theta) = \big[e^{-\jmath \frac{2\pi f}{c} (d_{k,1}-d_{k})},\cdots, e^{-\jmath \frac{2\pi f}{c} (d_{k,M}-d_{k})}\big]^{T}.
\end{equation}
Additionally, $g_{k}$ denotes the complex channel gain satisfying $|g_{k}|=\frac{c}{4\pi f d_{k}}$ ($1\leq k\leq K$), and $f$ denotes the operation frequency of the considered network.
On the other hand, due to the fact that near-field communications usually operate in the high-frequency band, the electromagnetic waves essentially propagate in straight pathes with poor bypassing capabilities and few scattering pathes. Thus, we adopt the Saleh-Valenzuela model for the NLoS channels \cite{X.Yu_MIMO_Hybrid}, which can be modeled as
\begin{equation}
\label{4}
\mathbf{h}_{\text{ID},k}^{\text{NLoS}} = \sum_{c=1}^{C}
g_{c,k}\mathbf{a}_{c}(d_{c},\theta_{c}),
\end{equation}
where $\mathbf{a}_{c}(d_{c},\theta_{c})=\big[e^{-\jmath \frac{2\pi f}{c} (d_{c,1}-d_{c})},\cdots, e^{-\jmath \frac{2\pi f}{c} (d_{c,M}-d_{c})}\big]^{T}$ denotes the array response vector from the BS to the $c$-th cluster. $g_{c,k}$ denotes the corresponding complex channel gain of BS$\rightarrow$scatter $c\rightarrow\text{ID}_{k}$ link, $d_{c}$ and $\theta_{c}$ denote distance and angle of the $c$-th cluster with respect to the original of the coordinate system. Also, $d_{c,k}=\|\mathbf{r}_k - \mathbf{s}_c\|$ denotes the distance from $c$-th cluster to the $\text{ID}_{k}$, where $\mathbf{s}_c=(d_{c}\cos\theta_{c},d_{c}\sin\theta_{c})$ denotes the coordinate of the $c$-th cluster. Accordingly, the near-field communication channel between the BS and the $\text{ID}_{k}$ is given by 
\begin{equation}
\label{5}
\mathbf{h}_{\text{ID},k}=\mathbf{h}_{\text{ID},k}^{\text{LoS}}+\mathbf{h}_{\text{ID},k}^{\text{NLoS}}.
\end{equation}
Similarly, the near-field energy charging between the BS and $\text{EH}_{l}$ is given by $\mathbf{h}_{\text{EH},l}=g_{l}\mathbf{b}(d_{l},\theta_{l})+\sqrt{\frac{1}{C}}\sum_{c=1}^{C}g_{c,l}\mathbf{a}_{c}(d_{c},\theta_{c})$, where $\mathbf{b}(d_{l},\theta_{l})$ is the array response vector from the BS to $\text{EH}_{l}$, $g_{l}$ and $g_{c,l}$ denote the corresponding complex channel gains. To obtain useful insights and characterize the fundamental performance upper bound of the considered near-field SWIPT system, it is assumed that the perfect CSI is available at the BS.

\subsection{Signal Model}
At the baseband digital module, $K$ data streams $\{s_{\text{ID},1},\cdots,s_{\text{ID},K}\}$ are modified via $K$ independent digital beamformers $\{\mathbf{w}_{1},\cdots,\mathbf{w}_{K}\}$, where $s_{\text{ID},k}\in\mathbb{C}$ denotes the information-bearing signal for $\text{ID}_{k}$, and $\mathbf{w}_{k}$ denotes the baseband digital beamforming vector for $\text{ID}_{k}$. Considering the fact that the energy harvesting is unaffected by the transmitted symbol contents, a common energy-carrying signal $s_{\text{EH}}$ is utilized as the input to the baseband digital module, which is a pseudo-random energy-carrying signal without being embedded any information. The output of the baseband digital module will be manipulated by the analog phase shifters, and then emitted to the antennas for wireless information and power transfer. Accordingly, the data/energy signal received at the IDs and EHs can be expressed as 
\begin{align}
\label{6}
\mathbf{x} =\sum_{k=1}^{K}\mathbf{P}\mathbf{w}_{k}s_{\text{ID},k}+
\sum_{l=1}^{\bar{L}}\mathbf{P}\mathbf{v}_{\bar{l}}s_{\text{EH}}.
\end{align}
To have the maximum flexibility for SWIPT beamforming design, we assume a variable number of energy beams, denoted by $\bar{L}$, which satisfies $0\leq \bar{L}\leq M_{\text{RF}}-K$. Thus, $\mathbf{v}_{\bar{l}}$ denotes the $\bar{l}$-th dedicated energy beamformer for signal $s_{\text{EH}}$. Note that $s_{\text{ID},k}$ and $s_{\text{EH}}$ satisfy $\mathbb{E}(|s_{\text{ID},k}|^{2})=\mathbb{E}(|s_{\text{EH}}|^{2})=1$, and each entry of analog beamformer should satisfy the following unit-modulus constraint, i.e.,
\begin{equation}
\label{7}
|\mathbf{P}^{[i,j]}|=1.
\end{equation}

Then, the received signal at the $\text{ID}_{k}$ and $\text{EH}_{l}$ can be expressed as
\begin{align}
\label{8}
y_{\text{ID},k}=\mathbf{h}_{\text{ID},k}^{H}\mathbf{x}+n_{\text{ID},k},\\
\label{9}
y_{\text{EH},l}=\mathbf{h}_{\text{EH},l}^{H}\mathbf{x}+n_{\text{EH},l},
\end{align}
where $n_{\text{ID},k},n_{\text{EH},l}\sim\mathcal{CN}(0,\sigma^{2})$ denotes the additive white Gaussian noise (AWGN) at the $\text{ID}_{k}$ and $\text{EH}_{l}$, respectively. Then, the achievable communication rate of the $\text{ID}_{k}$ is given by
\begin{align}
\label{10}
C_{k}=\log_{2}\left(1+\frac{|\mathbf{h}_{\text{ID},k}^{H}\mathbf{P}\mathbf{w}_{k}|^{2}}
{\sum_{i\neq k}^{K}|\mathbf{h}_{\text{ID},k}^{H}\mathbf{P}\mathbf{w}_{i}|^{2}+\sum_{\bar{l}=1}^{\bar{L}}
|\mathbf{h}_{\text{ID},k}^{H}\mathbf{P}\mathbf{v}_{\bar{l}}|^{2}+\sigma^{2}}\right).
\end{align}
Recalling the fact that the EHs can collect the energy of all electromagnetic waves in wireless channels, the received RF power at the $\text{EH}_{l}$ is given by
\begin{align}
\label{11}
Q_{l}\!=\!\eta\!\left[\sum_{k=1}^{K}|\mathbf{h}_{\text{EH},l}^{H}\mathbf{P}\mathbf{w}_{k}|^{2}+
\sum_{\bar{l}=1}^{\bar{L}}
|\mathbf{h}_{\text{EH},l}^{H}\mathbf{P}\mathbf{v}_{\bar{l}}|^{2}\right],
\end{align}
where $\eta$ denotes the energy harvesting efficiency.

\subsection{Problem Formulation}
Observe \eqref{10} and \eqref{11}, it knows that the dedicated energy beams cause the harmful interference to the IDs, while the information beams can be collected to benefit the energy harvest. Thus, a fundamental issue naturally appears, i.e., \textit{can we achieve the near-field SWIPT only using the information beams?} To address this problem, we aim to minimize the total power consumption at the BS by the joint optimization of analog beamformer, baseband digital information/energy beamformers, and the number of dedicated energy beams, subject to the individual QoS constraints of IDs and the energy harvesting constraints of EHs. As such, the problem is formulated as
\begin{subequations}
\begin{align}
\label{12a} (\text{P}1)\quad \min\limits_{\bar{L},\mathbf{P},\mathbf{w}_{k},\mathbf{v}_{\bar{l}}}\quad & \sum_{k=1}^{K}\|\mathbf{P}\mathbf{w}_{k}\|^{2}+
\sum_{\bar{l}=1}^{\bar{L}}\|\mathbf{P}\mathbf{v}_{\bar{l}}\|^{2}\\
\label{12b}\text{s.t.} \quad & C_{k}\geq \bar{C}_{k},\quad k\in \mathcal{K}_{\text{ID}},\\
\label{12c}  &Q_{l} \geq \bar{Q}_{l},\quad l\in \mathcal{K}_{\text{EH}},\\
\label{12d}  &|\mathbf{P}^{[i,j]}|=1, \quad 1\leq i\leq M, \quad 1\leq j\leq M_{\text{RF}},
\end{align}
\end{subequations}
where $\bar{C}_{k}$ and $\bar{Q}_{l}$ denote the QoS and RF receive power requirements of the $\text{ID}_{k}$ and the $\text{EH}_{l}$, while \eqref{12d} denotes the unit-modulus constraint of the analog beamformer. However, the problem (P1) is intractable to optimally solve due to uncertainty of number of dedicated energy beams and the high coupling between the analog beamformer and the baseband digital beamformers. Although by defining $\mathbf{W}_{k}=\mathbf{w}_{k}\mathbf{w}_{k}^{H}$, $\mathbf{V}_{\bar{l}}=\mathbf{v}_{\bar{l}}\mathbf{v}_{\bar{l}}^{H}$, and $\mathbf{V}=\sum_{\bar{l}=1}^{\bar{L}}\mathbf{V}_{\bar{l}}$, the SDR technique can be employed to optimize the baseband digital beamformers regardless of the uncertain $\bar{L}$, the resultant problem with $K+1$ matrix variables $\{\mathbf{W}_{1},\cdots,\mathbf{W}_{K},\mathbf{V}\}$ and $K+L$ constraints does not satisfy the Shapiro-Barvinok-Pataki rank-one condition \cite[eq. (32)]{Z.Luo_complexity}, where the tightness of the use of SDR cannot be guaranteed.

\section{Do We Need Dedicated Energy Beams in Near-field SWIPT?}\label{Section_3}
To eliminate the uncertainty caused by the number of dedicated energy beams, this section investigates whether the dedicated energy beams are necessary to the near-field SWIPT. We propose to employ the SDR technique to optimally design the baseband digital beamformers under a given analog beamformer. A generic rank-one solution construction method is proposed to guarantee the tightness of leveraging SDR for the considered problem. Based on the structure of rank-one solutions, the rank-zero property of the dedicated energy beam matrix is demonstrated.


\subsection{Problem Simplification}

The most straightforward way to address whether the dedicated energy beams are required for near-field SWIPT is to optimally solve the problem (P1). Denote the optimal value of the objective function of problem (P1) as $\mathfrak{O}(\mathbf{P}^{*},\mathbf{w}_{k}^{*},\mathbf{v}^{*})$, where $\{\mathbf{P}^{*},\mathbf{w}_{k}^{*},\mathbf{v}^{*}\}$ are the globally optimal solutions. We can check whether the equation $\mathbf{v}^{*}=\mathbf{0}$ holds to judge whether the dedicated energy beam is preferred. However, due to the non-convex unit-modulus constraints of analog beamformer, it is challenging to obtain the globally optimal solutions for problem (P1). To circumvent the effects of the non-convexity caused by the analog beamformer, we consider optimally optimizing the baseband digital beamformers under a given analog beamformer to investigate the necessity/non-necessity of the dedicated energy beams. It is easily to verify $\mathfrak{O}(\mathbf{P}^{*},\mathbf{w}_{k}^{*},\mathbf{v}^{*})\leq
\mathfrak{O}(\mathbf{P}^{*},\mathbf{w}_{k}^{*},\mathbf{0})\Leftrightarrow
\mathfrak{O}(\mathbf{P}^{\text{g}},\mathbf{w}_{k}^{*},\mathbf{v}^{*})\leq
\mathfrak{O}(\mathbf{P}^{\text{g}},\mathbf{w}_{k}^{*},\mathbf{0})$, which always holds since the operation of $\mathbf{v}^{*}=\mathbf{0}$ does not increase the feasible region of the original problem. In this case, the problem (P1) can be reduced to
\begin{subequations}
\begin{align}
\label{13a} (\text{P}1.1)\quad \min\limits_{\bar{L},\mathbf{w}_{k},\mathbf{v}_{\bar{l}}}\quad & \sum_{k=1}^{K}\|\mathbf{P}^{\text{g}}\mathbf{w}_{k}\|^{2}+
\sum_{\bar{l}=1}^{\bar{L}}\|\mathbf{P}^{\text{g}}\mathbf{v}_{\bar{l}}\|^{2}\\
\label{13b}\text{s.t.} \quad & C_{k}\geq \bar{C}_{k},\quad k\in \mathcal{K}_{\text{ID}},\\
\label{13c}  &Q_{l} \geq \bar{Q}_{l},\quad l\in \mathcal{K}_{\text{EH}}.
\end{align}
\end{subequations}

\subsection{Proposed Optimal Solutions}
To remove the uncertainty brought by $\bar{L}$, let $\mathbf{W}_{k}=\mathbf{w}_{k}\mathbf{w}_{k}^{H}$, $\mathbf{V}_{i}=\mathbf{v}_{i}\mathbf{v}_{i}^{H}$, $\mathbf{V}=\sum_{i=1}^{\bar{L}}\mathbf{V}_{i}$, and $\mathbf{\bar{P}}^{\text{g}}=(\mathbf{P}^{\text{g}})^{H}\mathbf{P}^{\text{g}}$. Hence, the problem (P1.1) can be reformulated as
\begin{subequations}
\begin{align}
\label{14a} (\text{P}1.2)\quad \min\limits_{\mathbf{W}_{k},\mathbf{V}}\quad & \sum_{k=1}^{K}\text{Tr}(\mathbf{\bar{P}}^{\text{g}}\mathbf{W}_{k})+\text{Tr}(\mathbf{\bar{P}}^{\text{g}}\mathbf{V})\\
\label{14b}\text{s.t.} \quad & \frac{\text{Tr}(\mathbf{\bar{H}}_{\text{ID},k}\mathbf{W}_{k})}{\bar{\gamma}_{k}}-
\sum_{i\neq k}^{K}\text{Tr}(\mathbf{\bar{H}}_{\text{ID},k}\mathbf{W}_{i})-\text{Tr}(\mathbf{\bar{H}}_{\text{ID},k}\mathbf{V})\geq \sigma^{2}, \ \  k\in \mathcal{K}_{\text{ID}},\\
\label{14c}  &\sum_{k=1}^{K}\text{Tr}(\mathbf{\bar{H}}_{\text{EH},l}\mathbf{W}_{k})+
\text{Tr}(\mathbf{\bar{H}}_{\text{EH},l}\mathbf{V}) \geq \frac{\bar{Q}_{l}}{\eta},\ \  l\in \mathcal{K}_{\text{EH}},\\
\label{14d}  &\mathbf{W}_{k}\succeq\mathbf{0},\ \ \text{rank}(\mathbf{W}_{k}) = 1,\ \   k\in \mathcal{K}_{\text{ID}},\\
\label{14e}  & \mathbf{V}\succeq\mathbf{0},
\end{align}
\end{subequations}
where $\bar{\gamma}_{k}=2^{\bar{C}_{k}}-1$, $\mathbf{\bar{H}}_{\text{ID},k}=\mathbf{\bar{h}}_{\text{ID},k}\mathbf{\bar{h}}_{\text{ID},k}^{H}$, $\mathbf{\bar{h}}_{\text{ID},k}=(\mathbf{P}^{\text{g}})^{H}\mathbf{h}_{\text{ID},k}$, $\mathbf{\bar{H}}_{\text{EH},l}=\mathbf{\bar{h}}_{\text{EH},l}\mathbf{\bar{h}}_{\text{EH},l}^{H}$, and $\mathbf{\bar{h}}_{\text{EH},l}=(\mathbf{P}^{\text{g}})^{H}\mathbf{h}_{\text{EH},l}$. Due to the non-convex rank-one constraint in \eqref{12d}, problem (P1.2) cannot be directly solved. To deal with this non-convex problem, we consider ignoring the rank-one constraint in problem (P1.2), and adopt the SDR technique to optimally solved the rank-one relaxed problem. However, since we neglect the rank-one constraint when solving the problem (P1.2), the obtained solutions may not satisfy the rank-one requirements, i.e., $\text{rank}(\mathbf{W}_{k}^{*})\geq1$ and $\text{rank}(\mathbf{V}^{*})\geq1$. To handle this issue, we consider constructing the optimal rank-one solutions of the problem (P1.2), which is given in Proposition \ref{Proposition_1}.
\begin{proposition}\label{Proposition_1}
    Assume that $\{\mathbf{W}_{k}^{*},\mathbf{V}^{*}\}$ are the optimal solutions of the problem (P1.2) with relaxed rank-one constraints. Then, we can always construct the rank-one solutions $\{\mathbf{\hat{W}}_{k}\}$ and $\{\mathbf{\hat{V}}\}$ that achieve the same performance as $\{\mathbf{W}_{k}^{*},\mathbf{V}^{*}\}$, which are given by
    \begin{gather}
    \label{15}
    \mathbf{\hat{w}}_{k} = (\mathbf{\bar{h}}_{\text{ID},k}^{H}\mathbf{W}_{k}^{*}\mathbf{\bar{h}}_{\text{ID},k})^{-\frac{1}{2}}\mathbf{W}_{k}^{*}
    \mathbf{\bar{h}}_{\text{ID},k}, \quad \mathbf{\hat{W}}_{k} =\mathbf{\hat{w}}_{k}(\mathbf{\hat{w}}_{k})^{H},\\ \label{16}
    \mathbf{\hat{V}} = \sum_{k=1}^{K}\mathbf{W}_{k}^{*}+\mathbf{V}^{*}-\sum_{k=1}^{K}\mathbf{\hat{W}}_{k}.
    \end{gather}
    Then, the dedicated energy beam vectors can be decomposed as $\mathbf{\hat{V}}=
    \sum_{\bar{l}=1}^{\text{rank}(\mathbf{\hat{V}})}\mathbf{\hat{v}}_{\bar{l}}(\mathbf{\hat{v}}_{\bar{l}})^{H}$ via the eigenvalue decomposition (EVD).
\end{proposition}
\begin{IEEEproof}
See Appendix A.
\end{IEEEproof}
\begin{remark}\label{Remark_1}
    From Proposition \ref{Proposition_1}, we can observe following two insights: 1) in the case of using dedicated energy beams, the SDR method is always tight for the problem (P1.2), and 2) the number of required energy beams is determined by the rank of $\mathbf{\hat{V}}$, i.e., $\bar{L}=\text{rank}(\mathbf{\hat{V}})$. It implies that by checking if $\text{rank}(\mathbf{\hat{V}})=0$ always holds, we can determine whether the dedicated energy beams are required.
\end{remark}



However, due to the uncertainty of the rank($\mathbf{W}^{*}$) and rank($\mathbf{V}^{*}$), it is difficult to analytically determine rank($\mathbf{\hat{V}}$). To tackle this issue, we focus on the proof of $\text{rank}(\mathbf{\hat{V}})=0$. Accordingly, Proposition \ref{Proposition_2} is introduced below.
\begin{proposition}\label{Proposition_2}
    Given the optimal solutions $\{\mathbf{\hat{W}}_{k},\mathbf{\hat{V}}\}$ of the problem (P1.2), we can always construct the other optimal solutions $\{\mathbf{\tilde{W}}_{k}\}$ without dedicated energy beams, which are given by
    \begin{gather}
    \label{17}
    \mathbf{\tilde{W}}_{k} = \mathbf{\hat{W}}_{k}+\alpha_{k}\mathbf{\hat{V}},
    \end{gather}
    where the arbitrary real weighted coefficients satisfy $\sum_{k=1}^{K}\alpha_{k}=1$ and $\alpha_{k}\geq 0$.
\end{proposition}
\begin{IEEEproof}
See Appendix B.
\end{IEEEproof}

With the results of Proposition \ref{Proposition_1} and Proposition \ref{Proposition_2}, it can be obtained that when omitting the rank-one constraints of the information beams, i.e., $\text{rank}(\mathbf{\tilde{W}}_{k})\geq1$, removing dedicated energy beam will not degrade the optimal performance of the considered network. Unfortunately, this conclusion may not applicable to the original optimization problem, which requires the rank-one solution. However, in the following proposition, we show that we can always construct a optimal rank-one solution from the high-rank solution obtained by SDR. To gain more insights, Proposition \ref{Proposition_3} is introduced.

\begin{proposition}\label{Proposition_3}
    With the optimal solutions $\{\mathbf{\tilde{W}}_{k}\}$ obtained in Proposition \ref{Proposition_2}, the rank-one solutions $\{\mathbf{\breve{W}}_{k}=\mathbf{\breve{w}}_{k}(\mathbf{\breve{w}}_{k})^{H}\}$ can be always constructed to achieve the same performance of $\{\mathbf{\tilde{W}}_{k}\}$ while satisfying all the constraints of the problem (P1.2).
\end{proposition}
\begin{IEEEproof}
     By substituting the optimal solutions $\{\mathbf{\tilde{W}}_{k}\}$ into problem (P1.2), we rewrite it as following form.
\begin{subequations}
\begin{align}
\label{18a} (\text{P}1.3)\quad \min\limits_{\mathbf{\tilde{W}}_{k}}\quad & \sum_{k=1}^{K}\text{Tr}(\mathbf{\bar{P}}^{\text{g}}\mathbf{\tilde{W}}_{k})\\
\label{18b}\text{s.t.} \quad & \sum_{k=1}^{K}\text{Tr}(\mathbf{\bar{A}}_{n,k}\mathbf{\tilde{W}}_{k})\geq \bar{a}_{n}, \ \ 1\leq n\leq K+L,\\
\label{18c}  &\mathbf{\tilde{W}}_{k}\succeq\mathbf{0},\ \   k\in \mathcal{K}_{\text{ID}}.
\end{align}
\end{subequations}
Thereinto, $\mathbf{\bar{A}}_{n,k}$ and $\bar{a}_{n}$ are defined by
\begin{align}\label{19}
   \mathbf{\bar{A}}_{n,k}=\begin{cases} \frac{\mathbf{\bar{H}}_{\text{ID},n}}{\bar{\gamma}_{n}},\ \  \quad \text{if}\  k=n,\\
   -\mathbf{\bar{H}}_{\text{ID},n}, \ \   \text{if}\   k\neq n,
   \end{cases} \quad \text{and}\qquad
   \bar{a}_{n}= \sigma^{2},\quad \text{if}\  1\leq n\leq K,
\end{align}
\begin{align}\label{20}
   \mathbf{\bar{A}}_{n,k}= \mathbf{\bar{H}}_{\text{EH},n-K}\quad \text{and}\qquad
   \bar{a}_{n}= \frac{\bar{Q}_{n-K}}{\eta}, \quad \text{if}\  K+1\leq n\leq K+L,
\end{align}
Considering the fact that the problem (P1.3) is a standard SDP problem, the strong-duality of problem (P1.3) always holds \cite[Th. 1.7.1, 4]{strong_duality}. Thus, the dual problem of the problem (P1.3) is given by
\begin{subequations}
\begin{align}
\label{21a} (\text{P}1.4)\quad \max\limits_{y_{n}}\quad & \sum_{n=1}^{K+L}y_{n}a_{n}\\
\label{21b} \text{s.t.} \quad &\mathbf{\bar{P}}^{\text{g}}-\sum_{n=1}^{K+L}y_{n}\mathbf{\bar{A}}_{n,k} \succeq \mathbf{0}, \ \ k\in \mathcal{K}_{\text{ID}},\\
\label{21c}   &y_{n}\geq 0,\ \   k\in \mathcal{K}_{\text{ID}}.
\end{align}
\end{subequations}
where $y_{n}$ is the optimization variable of the dual problem. According to the complementarity conditions, it is known that the globally optimal solutions $\{\mathbf{\tilde{W}}_{k}\}$ of problem (P1.3) must hold that
\begin{align}\label{22}
   \text{Tr}\left(\mathbf{\tilde{W}}_{k}\left(\mathbf{\bar{P}}^{\text{g}}-
   \sum_{n=1}^{K+L}y_{n}\mathbf{\bar{A}}_{n,k}\right)\right) = 0, \quad k\in \mathcal{K}_{\text{ID}},
\end{align}
By assuming $\text{rank}(\mathbf{\tilde{W}}_{k})=b_{k}>1$, we have $\mathbf{\tilde{W}}_{k}=\sum_{\iota=1}^{b_{k}}\mathbf{\tilde{w}}_{k,\iota}\mathbf{\tilde{w}}_{k,\iota}^{H}
=\mathbf{T}_{k}\mathbf{T}_{k}^{H}$ with $\mathbf{T}_{k}=[\mathbf{\tilde{w}}_{k,\iota},\cdots,\mathbf{\tilde{w}}_{k,b_{k}}]\in\mathbb{C}^{M_{\text{RF}}\times b_{k}}$. Then, we construct a rank-reduction Hermitian matrix $\mathbf{M}_{k}\in\mathbb{C}^{b_{k}\times b_{k}}$, which lies in the left null space of
$\mathbf{\bar{B}}_{k}\triangleq\left[\mathbf{T}_{k}^{H}\mathbf{\bar{A}}_{1,k}\mathbf{T}_{k},
\cdots,\mathbf{T}_{k}^{H}\mathbf{\bar{A}}_{K+L,k}\mathbf{T}_{k}\right]\in\mathbb{C}^{b_{k}\times (K+L)b_{k}}$, i.e.,
\begin{align}\label{23}
   \text{Tr}(\mathbf{\bar{B}}_{k}^{H}\mathbf{\bar{M}}_{k}) = \mathbf{0}.
\end{align}
Checking the structure of $\mathbf{\bar{B}}_{k}$, it is easily to find that: 1) $\text{rank}(\mathbf{\bar{B}}_{k})\leq b_{k}$; and 2) each sub-matrix block $\mathbf{T}_{k}^{H}\mathbf{\bar{A}}_{n,k}\mathbf{T}_{k}$ is a $b_{k}$-dimension rank-one matrix. From the first observation, we know that there are only $b_{k}$ columns in $\mathbf{\bar{B}}_{k}$ are linearly independent. However, the second observation indicates that the linear independence cannot happen inside each sub-matrix block $\mathbf{T}_{k}^{H}\mathbf{\bar{A}}_{n,k}\mathbf{T}_{k}$. Therefore, it readily knows that only $b_{k}$ sub-matrix blocks $\{\mathbf{T}_{k}^{H}\mathbf{\bar{A}}_{\bar{n},k}\mathbf{T}_{k}\}$ ($\bar{n} \in \mathcal{N}_{k} \triangleq\{1\leq \bar{n}\leq b_{k}\}$) are linearly independent, while all the remaining $\mathbf{T}_{k}^{H}\mathbf{\bar{A}}_{\bar{m},k}\mathbf{T}_{k}$ ($\bar{m} \neq\bar{n}\in \mathcal{N}_{k} $) are linearly dependent. This indicates that any
$\mathbf{T}_{k}^{H}\mathbf{\bar{A}}_{\bar{m},k}\mathbf{T}_{k}$ can be expressed as the linear combination form of $\mathbf{T}_{k}^{H}\mathbf{\bar{A}}_{\bar{n},k}\mathbf{T}_{k}$ ($\bar{n} \in \mathcal{N}_{k}$). On this basis, we can equivalently transform \eqref{23} to the following linear equations:
\begin{align}\label{24}
\text{Tr}(\mathbf{T}_{k}^{H}\mathbf{\bar{A}}_{\bar{n},k}\mathbf{T}_{k}\mathbf{M}_{k})=0, \quad \bar{n} \in \mathcal{N}_{k},
\end{align}
which has $b_{k}$ equations and $b_{k}^2$ real-valued variables. To guarantee the non-zero solution $\mathbf{M}_{k}$, we should assume $b_{k}^2\geq b_{k}$ hold. In this case, we can construct a matrix $\mathbf{\bar{M}}_{k}=\frac{1}{\delta_{k}}\mathbf{M}_{k}$, where $\delta_{k}$ denotes the maximal eigenvalue of $\mathbf{M}_{k}$. Accordingly, the rank-reduced solution $\mathbf{\breve{W}}_{k}$ can be construct by
\begin{align}\label{25}
\mathbf{\breve{W}}_{k}=\mathbf{T}_{k}(\mathbf{I}-\mathbf{\bar{M}}_{k})\mathbf{T}_{k}^{H},
\end{align}
which is a semidefinite matrix due to the fact of $\mathbf{I}-\mathbf{\bar{M}}_{k}\succeq\mathbf{0}$. Also, since $\text{Tr}(\mathbf{T}_{k}^{H}\mathbf{\bar{A}}_{n,k}\mathbf{T}_{k}\mathbf{M}_{k})=0$ satisfies all $\mathbf{\bar{A}}_{n,k}$, it is easily to verify that
\begin{align}\label{26}
\text{Tr}(\mathbf{\bar{A}}_{n,k}\mathbf{\breve{W}}_{k})=\text{Tr}(\mathbf{\bar{A}}_{n,k}\mathbf{\tilde{W}}_{k})
-\text{Tr}(\mathbf{T}_{k}^{H}\mathbf{\bar{A}}_{n,k}\mathbf{T}_{k}\mathbf{M}_{k})=\text{Tr}(\mathbf{\bar{A}}_{n,k}\mathbf{\tilde{W}}_{k}),
\end{align}
which proves that rank-reduced solutions $\{\mathbf{\breve{W}}_{k}\}$ satisfy the constraints of the problem (P1.3). With the complementarity conditions in \eqref{22}, we know $\mathbf{T}_{k}^{H}\left(\mathbf{\bar{P}}^{\text{g}}-
   \sum_{n=1}^{K+L}y_{n}\mathbf{\bar{A}}_{n,k}\right)\mathbf{T}_{k} =\mathbf{0}$. Hence, the rank-reduced solution $\mathbf{\breve{W}}_{k}$ must satisfy the complementarity conditions, i.e.,
\begin{align} \nonumber
&\text{Tr}\left(\mathbf{T}_{k}(\mathbf{I}-\mathbf{\bar{M}}_{k})\mathbf{T}_{k}^{H}\left(\mathbf{\bar{P}}^{\text{g}}-
   \sum_{n=1}^{K+L}y_{n}\mathbf{\bar{A}}_{n,k}\right)\right)=\\ \label{27}
&\qquad\qquad\qquad \text{Tr}\left((\mathbf{I}-\mathbf{\bar{M}}_{k})\mathbf{T}_{k}^{H}\left(\mathbf{\bar{P}}^{\text{g}}-
   \sum_{n=1}^{K+L}y_{n}\mathbf{\bar{A}}_{n,k}\right)\mathbf{T}_{k}\right) = 0,
\end{align}
which proves the optimality of $\mathbf{\breve{W}}_{k}$. After above procedure, the rank of $\mathbf{\bar{W}}_{k}$ decreases by at least one, i.e., $\text{rank}(\mathbf{\breve{W}}_{k})\leq \text{rank}(\mathbf{\bar{W}}_{k})-1$ \cite{Y.Huang_SSDP}. Therefore, we can repeat this process until $b_{k}^2\leq b_{k}$. When $b_{k}\leq1$, we cannot construct a non-zero $\mathbf{M}_{k}$ anymore, and the rank-one solutions are obtained. This completes the proof.
\end{IEEEproof}


Following the result of Proposition \ref{Proposition_3}, we can obtain an important insight for dedicated energy beam design. To elaborate, it knows that the dedicated energy beams can be incorporated into the information beams. Thus, we can rewrite the rank-one information beam $\mathbf{\breve{W}}_{k}$ as
\begin{align}\label{28}
\mathbf{\breve{W}}_{k}= \mathbf{\dot{W}}_{k}+\alpha_{k}\mathbf{\hat{V}},
\end{align}
where $\mathbf{\dot{W}}_{k}$ is a non-zero matrix as the information beam is always required. Thus, we have
\begin{align}\label{29}
\text{rank}(\mathbf{\breve{W}}_{k})=\text{rank}(\mathbf{\dot{W}}_{k}+\alpha_{k}\mathbf{\hat{V}})=1, \quad \text{for any} \ k\in \mathcal{K}_{\text{ID}}.
\end{align}
To guarantee the above equation always held, it readily verifies that the dedicated energy beams obtained in Proposition \ref{Proposition_1} must satisfy one of conditions: 1) if there exist at least two $\{\mathbf{\dot{W}}_{k},\mathbf{\dot{W}}_{\tilde{k}}\}$ ($k\neq\tilde{k}$) that are linearly independent, $\mathbf{\hat{V}}=\mathbf{0}$; and 2) if all the $\mathbf{\dot{W}}_{k}$ ($k\in \mathcal{K}_{\text{ID}}$) are linearly dependent, $\mathbf{\hat{V}}$ is also linearly dependent on $\mathbf{\dot{W}}_{k}$. For the former case, it knows that $\text{rank}(\mathbf{\hat{V}})=0$ without changing the objective function's value. For the latter case, since it has $\mathbf{\hat{V}}=\bar{\alpha}_{k}\mathbf{\dot{W}}_{k}$, the any non-zero dedicated energy beam can be fully incorporated into the information beams, which leaves the remaining dedicated energy beams satisfying $\text{rank}(\mathbf{\hat{V}})=0$. Based on the above discussion, it is concluded that under any feasible analog beamformer, the dedicated energy beams are not required, i.e., optimal $\bar{L}^{*}$ satisfies $\bar{L}^{*}=0$.

\section{Hybrid Beamforming Design}\label{Section_4}
In this section, we concentrate on the design of the hybrid beamforming for the near-field SWIPT network. Based on the conclusion that the dedicated energy beams are no needed, we focus on the information beams design. A PTL algorithm is developed, where the analog beamformer and baseband digital beamformers are derived in the closed-form expressions. Then, a two-stage hybrid beamforming strategy is proposed for further reducing the computational complexity of beamforming design. 

\subsection{Proposed PTL Algorithm}
With the definitions of $\mathbf{W}_{\text{I}}=[\mathbf{w}_{1},\cdots,\mathbf{w}_{K}]$, $\mathbf{W}_{k}=\mathbf{w}_{k}\mathbf{w}_{k}^{H}$, $\mathbf{H}_{\text{ID},k}=\mathbf{h}_{\text{ID},k}\mathbf{h}_{\text{ID},k}^{H}$ and $\mathbf{H}_{\text{EH},l}=\mathbf{h}_{\text{EH},l}\mathbf{h}_{\text{EH},l}^{H}$, we introduce an auxiliary variable of $\mathbf{U}_{\text{I}}$, which satisfy $\mathbf{U}_{\text{I}}=\mathbf{P}\mathbf{W}_{\text{I}}$. By integrating the above equality constraint into the objective function as the penalty term, the optimization problem (P1) can be reformulated as following augmented Lagrangian (AL) form.
\begin{subequations}
\begin{align}
\label{30a} (\text{P}2)\ \  \min\limits_{\mathbf{W}_{k},\mathbf{W}_{\text{I}},\mathbf{P},\mathbf{U}_{\text{I}}}\quad & \left\|\mathbf{U}_{\text{I}}\right\|_{F}^{2}+
\frac{1}{2\varrho}\left(\left\|\mathbf{P}\mathbf{W}_{\text{I}}-\mathbf{U}_{\text{I}}\right\|_{F}^{2}\right)\\
\label{30b}\text{s.t.} \quad & \frac{\text{Tr}(\mathbf{P}^{H}\mathbf{H}_{\text{ID},k}\mathbf{P}\mathbf{W}_{k})}{\bar{\gamma}_{k}}\!-\!
\sum_{i\neq k}^{K}\text{Tr}(\mathbf{P}^{H}\mathbf{H}_{\text{ID},k}\mathbf{P}\mathbf{W}_{i})\geq \sigma^{2},  k\in \mathcal{K}_{\text{ID}},\\
\label{30c}  &\sum_{k=1}^{K}\text{Tr}(\mathbf{P}^{H}\mathbf{H}_{\text{EH},l}\mathbf{P}\mathbf{W}_{k}) \geq \frac{\bar{Q}_{l}}{\eta},\ \  l\in \mathcal{K}_{\text{EH}},\\
\label{30d}  &\mathbf{W}_{k}\succeq\mathbf{0},\ \ \text{rank}(\mathbf{W}_{k}) = 1,\ \   k\in \mathcal{K}_{\text{ID}},\\
\label{30e}  & |\mathbf{P}^{[i,j]}|=1,  \quad 1\leq i\leq M, \quad 1\leq j\leq M_{\text{RF}},
\end{align}
\end{subequations}
where $\varrho>0$ is the penalty factor. Note that when $\varrho\rightarrow 0$ holds, the equality constraints $\mathbf{U}_{\text{I}}=\mathbf{P}\mathbf{W}_{\text{I}}$ can be guaranteed. However, initializing to a very small value $\varrho$ is actually impractical. This is because in this case, the penalty terms will predominate in the objective function \eqref{30a}, and the original objective function $\left\|\mathbf{U}_{\text{I}}\right\|_{F}^{2}$ is neglected. To tackle this issue, we adopt the penalty-based two-layer structure \cite{PDD}, where the optimization variables $\{\mathbf{W}_{k},\mathbf{W}_{\text{I}},\mathbf{P},\mathbf{U}_{\text{I}}\}$ are optimized in the inner layer under the given $\varrho$, while $\varrho$ is updated in the outer layer for guaranteeing the feasibility of the obtained solutions. To proceed, we further introduce the slack variables $\{\mathbf{p}_{k},\mathbf{q}_{l}\}$ to deal with the coupling between $\mathbf{W}_{k}$ and $\mathbf{P}$ in \eqref{30b} and \eqref{30c}, which satisfy $\mathbf{p}_{k}=\mathbf{h}_{\text{ID},k}^{H}\mathbf{U}_{\text{I}}$ and $\mathbf{q}_{l}=\mathbf{h}_{\text{EH},l}^{H}\mathbf{U}_{\text{I}}$. Accordingly, the problem (P2) can be rewritten as
\begin{subequations}
\begin{align}
\label{31a} (\text{P}2.1)\ \  \min\limits_{\mathbf{W}_{\text{I}},\mathbf{P},\mathbf{U}_{\text{I}},
\atop
\mathbf{p}_{k},\mathbf{q}_{l}}\quad & \left\|\mathbf{U}_{\text{I}}\right\|_{F}^{2}+
\frac{1}{2\varrho}\left(\left\|\mathbf{P}\mathbf{W}_{\text{I}}-\mathbf{U}_{\text{I}}\right\|_{F}^{2}\right)\\
\label{31b}\text{s.t.} \quad & \frac{(1+\bar{\gamma}_{k})|\mathbf{p}_{k}^{[k]}|^{2}}{\bar{\gamma}_{k}}\geq \|\mathbf{p}_{k}\|^{2}+\sigma^{2},  k\in \mathcal{K}_{\text{ID}},\\
\label{31c}  &\sum_{k=1}^{K}|\mathbf{q}_{l}^{[k]}|^{2} \geq \frac{\bar{Q}_{l}}{\eta},\ \  l\in \mathcal{K}_{\text{EH}},\\
\label{31d}  &\mathbf{p}_{k}=\mathbf{h}_{\text{ID},k}^{H}\mathbf{U}_{\text{I}},\quad
\mathbf{q}_{l}=\mathbf{h}_{\text{EH},l}^{H}\mathbf{U}_{\text{I}}, \quad k\in \mathcal{K}_{\text{ID}},\ l\in \mathcal{K}_{\text{EH}},\\
\label{31e}  &\eqref{30d},\eqref{30e},
\end{align}
\end{subequations}
Obviously, problem (P2.1) is still intractable to solve due to the existence of the quadratic terms in \eqref{31b} and \eqref{31c}. However, we can observe that the optimization variables are fully separated in the constraints, which motivates us to adopt BCD approach to solve the problem (P2.1) in the inner layer. More specifically, we partition the all the optimization variables into three blocks $\{\mathbf{U}_{\text{I}},\mathbf{p}_{k},\mathbf{q}_{l}\}$, $\{\mathbf{W}_{\text{I}}\}$ and $\{\mathbf{P}\}$, each of which are optimized while remaining the other blocks constant.

\subsubsection{Inner layer iteration: solve subproblem with respect to $\{\mathbf{U}_{\text{I}},\mathbf{p}_{k},\mathbf{q}_{l}\}$} Given fixed $\{\mathbf{W}_{\text{I}},\mathbf{P}\}$, problem (P2.1) is rewritten as
\begin{subequations}
\begin{align}
\label{32a} (\text{P}3.1)\ \  \min\limits_{\mathbf{U}_{\text{I}},
\mathbf{p}_{k},\mathbf{q}_{l}}\quad & \left\|\mathbf{U}_{\text{I}}\right\|_{F}^{2}+
\frac{1}{2\varrho}\left(\left\|\mathbf{P}\mathbf{W}_{\text{I}}-\mathbf{U}_{\text{I}}\right\|_{F}^{2}\right)\\
\label{32b}\text{s.t.} \quad & \frac{(1+\bar{\gamma}_{k})|\mathbf{p}_{k}^{[k]}|^{2}}{\bar{\gamma}_{k}}
\geq \|\mathbf{p}_{k}\|^{2}+\sigma^{2},  k\in \mathcal{K}_{\text{ID}},\\
\label{32c}  &\sum_{k=1}^{K}|\mathbf{q}_{l}^{[k]}|^{2}\geq \frac{\bar{Q}_{l}}{\eta},\ \  l\in \mathcal{K}_{\text{EH}},\\
\label{32d}  &\mathbf{p}_{k}=\mathbf{h}_{\text{ID},k}^{H}\mathbf{U}_{\text{I}},\quad
\mathbf{q}_{l}=\mathbf{h}_{\text{EH},l}^{H}\mathbf{U}_{\text{I}}, \quad k\in \mathcal{K}_{\text{ID}},\ l\in \mathcal{K}_{\text{EH}}.
\end{align}
\end{subequations}
To solve the non-convex constraint, the first-order Taylor expansion is adopted to construct the linear lower bound of quadratic term $|\mathbf{p}_{k}^{[k]}|^{2}$ as follows:
\begin{align}\label{33}
\varpi_{k}\triangleq2\Re\{(\mathbf{\dot{p}}_{k}^{[k]})^{H}\mathbf{p}_{k}^{[k]}\}-
|\mathbf{\dot{p}}_{k}^{[k]}|^2\leq\|\mathbf{p}_{k}^{[k]}\|^{2},
\end{align}
where $\mathbf{\dot{p}}_{k}$ denote the optimized result of $\mathbf{p}_{k}$ in the previous iteration, whose $k$-th element is $\mathbf{\dot{p}}_{k}^{[k]}$. Similarly, $|\mathbf{q}_{l}^{[k]}|^{2}$ can be approximated by $\omega_{l,k}\triangleq2\Re\{(\mathbf{\dot{q}}_{l}^{[k]})^{H}\mathbf{q}_{l}^{[k]}\}-
|\mathbf{\dot{q}}_{l}^{[k]}|^2$. Then, the problem (P3.1) can be efficiently solved by employing the SCA method, where the approximated problem at each iteration is given by
\begin{subequations}
\begin{align}
\label{34a} (\text{P}3.2)\ \  \min\limits_{\mathbf{U}_{\text{I}},
\mathbf{p}_{k},\mathbf{q}_{l}}\quad & \left\|\mathbf{U}_{\text{I}}\right\|_{F}^{2}+
\frac{1}{2\varrho}\left(\left\|\mathbf{P}\mathbf{W}_{\text{I}}-\mathbf{U}_{\text{I}}\right\|_{F}^{2}\right)\\
\label{34b}\text{s.t.} \quad & \frac{(1+\bar{\gamma}_{k})\varpi_{k}}{\bar{\gamma}_{k}}\geq \|\mathbf{p}_{k}\|^{2}+\sigma^{2},  k\in \mathcal{K}_{\text{ID}},\\
\label{34c}  &\sum_{k=1}^{K}\omega_{l,k}\geq \frac{\bar{Q}_{l}}{\eta},\ \  l\in \mathcal{K}_{\text{EH}},\\
\label{34d}  &\eqref{32d}.
\end{align}
\end{subequations}
The detailed algorithm for solving problem (P3.2) in summarized in \textbf{Algorithm \ref{SCA}}, where the analog beamformer $\mathbf{P}$ and baseband digital beamformers $\mathbf{W}_{\text{I}}$ are randomly initialized in the feasible region.
\begin{algorithm}[t]
    \caption{SCA algorithm.}
    \label{SCA}
    \begin{algorithmic}[1]
        \STATE{Initialize $\mathbf{P}$, $\mathbf{W}_{\text{I}}$, $\mathbf{\dot{p}}_{k}=\mathbf{\dot{q}}_{l}=[1,\cdots,1]$, and $m=1$. Set the convergence accuracy $\epsilon_{1}$.}
        \REPEAT
        \STATE{ update $\mathbf{U}_{\text{I}}$ by solving problem (P3.1).}
        \STATE{ update $\mathbf{\dot{p}}_{k}=\mathbf{p}_{k}$ and $\mathbf{\dot{q}}_{l}=\mathbf{q}_{l}$.}
        \STATE{ denote the objective value at $m$-th iteration as $g^{m}$ and set $m=m+1$.}
        \UNTIL{$|g^{m-1}-g^{m}|\leq \epsilon_{1}$}
    \end{algorithmic}
\end{algorithm}

\subsubsection{Inner layer iteration: solve subproblem with respect to $\{\mathbf{W}_{\text{I}}\}$} Given fixed $\{\mathbf{U}_{\text{I}},\mathbf{p}_{k},\mathbf{q}_{l},\mathbf{P}\}$, problem (P2.1) is reduced to $\min\limits_{\mathbf{W}_{\text{I}}}\  \left\|\mathbf{P}\mathbf{W}_{\text{I}}-\mathbf{U}_{\text{I}}\right\|_{F}^{2}$. It is a convex program, and can be optimally solved by the first-order optimality condition, where the optimal solution is given by
\begin{align}\label{35}
\mathbf{W}_{\text{I}}=(\mathbf{P}^{H}\mathbf{P})^{-1}\mathbf{P}^{H}\mathbf{U}_{\text{I}}.
\end{align}

\subsubsection{Inner layer iteration: solve subproblem with respect to $\{\mathbf{P}\}$} With fixed $\{\mathbf{U}_{\text{I}},\mathbf{p}_{k},\mathbf{q}_{l},\mathbf{W}_{\text{I}}\}$, the problem (P3.1) is converted to
\begin{subequations}
\begin{align}
\label{36a} (\text{P}3.3)\ \  \min\limits_{\mathbf{P}}\quad & \left\|\mathbf{P}\mathbf{W}_{\text{I}}-\mathbf{U}_{\text{I}}\right\|_{F}^{2}\\
\label{36b}\text{s.t.} \quad & \eqref{30e}.
\end{align}
\end{subequations}
To tackle the unit-modulus constraint \eqref{30e}, we rewrite problem (P3.3) as
\begin{subequations}
\begin{align}
\label{37a} (\text{P}3.4)\ \  \min\limits_{\mathbf{P}}\quad & \text{Tr}(\mathbf{P}^{H}\mathbf{P}\mathbf{Y})-2\Re(\text{Tr}(\mathbf{P}\mathbf{Z}))\\
\label{37b}\text{s.t.} \quad & \eqref{30e}.
\end{align}
\end{subequations}
where $\mathbf{Y}=\mathbf{W}\mathbf{W}^{H}$ and $\mathbf{Z}=\mathbf{U}_{\text{I}}\mathbf{W}^{H}$. It is easily observed that the element of $\mathbf{P}$ are separated in the constraint \eqref{30e}, which motivates us to adopt tht BCD method to optimize $\mathbf{P}$ in an element-wise manner. Specifically, we optimize $\mathbf{P}^{[i,j]}$ with remaining the other elements in $\mathbf{P}$ constant. As a result, the optimization problem for solving $\mathbf{P}^{[i,j]}$ is expressed as
\begin{subequations}
\begin{align}
\label{38a} (\text{P}3.5)\ \  \max\limits_{\mathbf{P}^{[i,j]}}\quad & \Re\left\{\psi_{i,j}\mathbf{P}^{[i,j]}\right\}+
\chi_{i,j}|\mathbf{P}^{[i,j]}|^{2}\\
\label{38b}\text{s.t.} \quad & \eqref{30e},
\end{align}
\end{subequations}
where $\psi_{i,j}$ is a complex constant coefficient, and $\chi_{i,j}$ is a real constant coefficient. With the fact of $|\mathbf{P}^{[i,j]}|^{2}=1$,  the optimal solution of problem (P3.4) is obtained by
\begin{align}\label{39}
\mathbf{P}^{[i,j]}=\frac{\psi_{i,j}}{\left|\psi_{i,j}\right|}.
\end{align}
To derive the expression of $\psi_{i,j}$, we have
\begin{align}\label{40}
\frac{\partial\mathcal{F}(\mathbf{P})}
{\partial\mathbf{P}^{[i,j]}}\bigg|_{\mathbf{P}^{[i,j]}=\mathbf{\dot{P}}^{[i,j]}}
=\frac{1}{2}(\chi_{i,j}\mathbf{\dot{P}}^{[i,j]}-\psi_{i,j})
=\frac{1}{2}(\mathbf{\dot{P}}\mathbf{Y}-\mathbf{Z})^{[i,j]},
\end{align}
where $\mathcal{F}(\mathbf{P})=\text{Tr}(\mathbf{P}^{H}\mathbf{P}\mathbf{Y})-2\Re(\text{Tr}(\mathbf{P}\mathbf{Z}))$, and $\mathbf{\dot{P}}$ is the optimized result of $\mathbf{P}$ in the previous iteration. By expanding $\mathbf{\dot{P}}\mathbf{Y}$, we can obtain $\chi_{i,j}\mathbf{\dot{P}}^{[i,j]}=\mathbf{\dot{P}}^{[i,j]}\mathbf{Y}^{[j,j]}$. Thus, the expression of $\psi_{i,j}$ is given by
\begin{align}\label{41}
\psi_{i,j}=\mathbf{Z}^{[j,j]}-(\mathbf{\dot{P}}\mathbf{X})^{[i,j]}+
\mathbf{\dot{P}}^{[i,j]}\mathbf{Y}^{[j,j]}.
\end{align}
By substituting \eqref{41} into \eqref{39}, the entries of $\mathbf{P}$ can be recursively updated, where at least the local minimizer of $\mathbf{P}$ is obtained \cite{S.Boyd}.

\subsubsection{Outer layer iteration}
To ensure that the obtained the solutions satisfy the equality constraints $\mathbf{U}_{\text{I}}=\mathbf{P}\mathbf{W}_{\text{I}}$, we consider gradually decreasing the penalty factor $\varrho$ in the outer layer,, i.e.,
\begin{align}\label{42}
\varrho=\frac{\varrho}{c'},
\end{align}
where $c'$ is a constant coefficient with satisfying $c'>1$. Note that when $\varrho<\epsilon_{3}\ll 1$, the obtained $\{\mathbf{P}, \mathbf{W}_{\text{I}}\}$ are feasible to the original problem (P2).

\begin{algorithm}[t]
    \caption{Penalty-based two-layer algorithm.}
    \label{PTL}
    \begin{algorithmic}[1]
        \STATE{Initialize $\mathbf{P}$, $\mathbf{W}_{\text{I}}$, $\mathbf{\dot{p}}_{k}$, $\mathbf{\dot{q}}_{l}$, $l=1$. Set the convergence accuracy $\epsilon_{2}$ and $\epsilon_{3}$.}
        \REPEAT
        \REPEAT
        \STATE{ update $\mathbf{U}_{\text{I}}$ by carrying out \textbf{Algorithm \ref{SCA}}.}
        \STATE{ update $\mathbf{W}_{\text{I}}$ according to \eqref{35}.}
        \STATE{ update $\mathbf{P}$ according to \eqref{39}.}
        \STATE{ calculate the objective value at $l$-th iteration as $f^{l}$, and set $l=l+1$.}
        \UNTIL{$|f^{l-1}-f^{l}|\leq \epsilon_{2}$.}
        \STATE{ $\varrho=\frac{\varrho}{c'}$.}
        \UNTIL{$\varrho\leq \epsilon_{3}$.}
    \end{algorithmic}
\end{algorithm}

The overall algorithm is summarized in \textbf{Algorithm \ref{PTL}}. Denoting the objective function value at $l$-th iteration of inner layer as the function $f(\mathbf{U}_{\text{I}}^{l},\mathbf{W}_{\text{I}}^{l},\mathbf{P}^{l})$, we have following inequality in the inner layer
\begin{align}
\label{43} f(\mathbf{U}_{\text{I}}^{l-1},\mathbf{W}_{\text{I}}^{l-1},\mathbf{P}^{l-1}) \overset{a}{\geq} f(\mathbf{U}_{\text{I}}^{l},\mathbf{W}_{\text{I}}^{l-1},\mathbf{P}^{l-1})\overset{b}{\geq}   f(\mathbf{U}_{\text{I}}^{l},\mathbf{W}_{\text{I}}^{l},\mathbf{P}^{l-1})\overset{c}{\geq}   f(\mathbf{U}_{\text{I}}^{l},\mathbf{W}_{\text{I}}^{l},\mathbf{P}^{l}).
\end{align}
Here, inequality sign $a$ holds as the optimal baseband digital $\mathbf{U}_{\text{I}}$ is obtained in \textbf{Algorithm \ref{SCA}} can be always guaranteed, the inequality sign $b$ holds as the optimal $\mathbf{W}_{\text{I}}$ is guaranteed at step $5$, and inequality sign $c$ holds because that at least the local minimizer of $\mathbf{P}$ is returned. Meanwhile, by gradually decreasing penalty factor, all the equality constraints can be satisfied. With the fact that optimal value is bounded, the stationary point solution of original problem can be found by the proposed PTL algorithm.

The main complexity of the proposed algorithm relies on solving the problem (P3.2) and the inverse operation in \eqref{35}. The problem (P3.2) is a second order cone programming (SOCP) program, which can be optimally solved by the interior-point method with complexity of $\mathcal{O}\big((MM_{\text{RF}}+K^2+KL)^{3.5}\big)$. Meanwhile, it is known that the complexity of inverse operation is given by $\mathcal{O}\big(M_{\text{RF}}^{3}\big)$. Hence, the overall complexity based on the interior-point method is given by $\mathcal{O}\Big(\log(\frac{1}{\epsilon_{3}})\\ \big[ \log(\frac{1}{\epsilon_{1}})(MM_{\text{RF}}+K^2+KL)^{3.5}
+M_{\text{RF}}^{3}\big]\Big)$.

\subsection{Low-complexity Hybrid Beamforming Design}
In the previous subsections, a PTL algorithm is proposed to jointly design the baseband digital beamformers and analog beamformer. However, in near-field communications, the BS is usually equipped with ELAA, which inevitably imposes the heavy computational demands on the BS. To tackle this issue, we propose a two-stage hybrid beamforming algorithm to reduce the computational complexity at the BS. To elaborate, since the NLoS links suffer the double path loss, the LoS channel is dominant in the considered network. Therefore, we consider designing the analog beamformer by maximizing the array gain at each ID/EH \cite{LDMA}. In this case, the analog precoding vector $\mathbf{p}_{j}$ is aligned to the array response vector of ID/EH, which is given by
\begin{align}
\label{44}
\mathbf{p}_{j}=\begin{cases}\mathbf{a}(d_{j},\theta_{j}), \quad &\text{if } 1\leq j\leq K, \\
\mathbf{b}(d_{j-K},\theta_{j-K}), \quad &\text{if } K+1\leq j\leq K+L,\\
\end{cases}
\end{align}
While for the $\mathbf{p}_{j}$ ($K+L\leq j\leq M_{\text{RF}}$), we consider exploiting them to enhance the array gain at all the EHs, which thus is designed by
\begin{align}
\label{45}
\mathbf{p}_{j}=\frac{1}{|\mathbf{c}|}\odot\mathbf{c}, \quad K+L\leq j\leq M_{\text{RF}},
\end{align}
where $\mathbf{c}=\sum_{l=1}^{L}\mathbf{b}(d_{l},\theta_{l})$. With above analog precorders, we can obtain the optimal baseband digital beamformers by solving the problem (P1.2). Specifically, if $\text{rank}(\mathbf{\tilde{W}}_{k})=1$ ($1\leq k\leq K$), the optimal baseband digital beamformers can be obtained by carrying out the EVD to $\mathbf{\tilde{W}}_{k}$; 2) if $\text{rank}(\mathbf{\tilde{W}}_{k})>1$ ($1\leq k\leq K$), we can adopt the rank-reduction method in Proposition \ref{Proposition_3} to find the optimal rank-one solutions based on high-rank solution $\mathbf{\tilde{W}}_{k}$. The main complexity of two-stage hybrid beamforming algorithm depends on solving SDP problem (P1.2), so the computational complexity by employing the interior-point method is given by $\mathcal{O}(M_{\text{RF}}^{3.5})$. Note that the complexity of the two-stage hybrid beamforming algorithm is independent of the number of the transmit antenna $M$, which thus is more efficient than the PTL algorithm in the practice.

\section{Numerical Results}\label{Section_5}
\begin{figure}[t]
\centering
\begin{minipage}[t]{0.45\textwidth}
\centering
\includegraphics[width=1\textwidth]{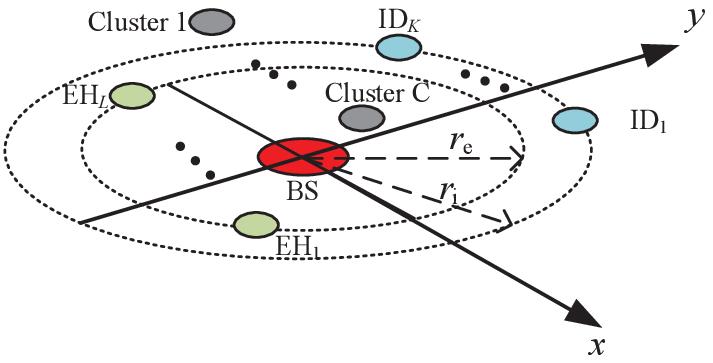}
\caption{Illustration of the simulation setup.}
\label{Fig_setup}
\end{minipage}\qquad
\begin{minipage}[t]{0.45\textwidth}
\centering
\includegraphics[width=1\textwidth]{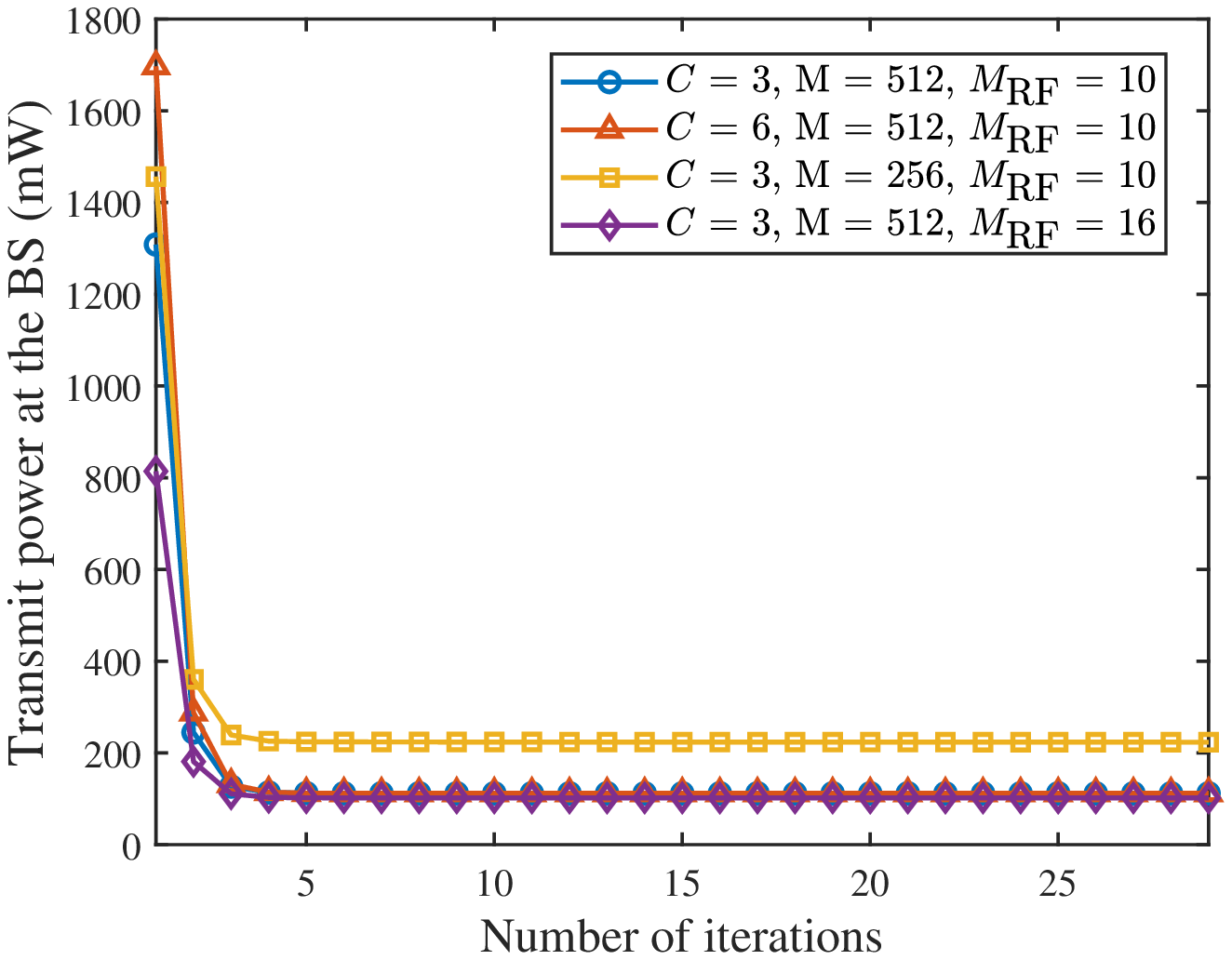}
\caption{Convergence performance with $K = 3$, $L = 3$, $\bar{C}=1$ bps/Hz, $\bar{Q}=20$ mW, and $r_{\text{i}}=r_{\text{e}}=20$ m.}
\label{Fig.2}
\end{minipage}
\end{figure}

This section numerically evaluates the performance of  the proposed algorithms. We consider a linear topology network setup as shown in Fig. \ref{Fig_setup}, where the BS is located in $(0, 0)$ meter (m). Without loss of generality, we assume that $\text{ID}_{k}$ and $\text{EH}_{l}$ are randomly located on the circles with radiuses $r_{\text{i}}$ m and $r_{\text{e}}$ m respectively, where the BS is located at the centres of these circles. The clusters are assumed to be randomly distributed in a circle with radius of $30$ m. The carrier frequency $f$ is 28 GHz, and the sub-wavelength antenna pitch is considered. For fairness, we assume all the IDs have the same QoS requirement, and all the EHs have the same RF receive power demand, i.e., $\bar{C}_{k}=\bar{C}$ and $\bar{Q}_{l}=\bar{Q}$. The common simulation parameters are listed in Table \ref{tab-2}. The other parameters are presented in the caption of each figure. Moreover, each numerical result is averaged over the 100 independent channel realizations.
\begin{table}[t]
\centering
    \caption{System parameters}\label{tab-2}
\begin{tabular}{|c|c|}
		\hline
        \tabincell{c}{Carrier frequency} & $f = 28$ GHz \\
		\hline
        \tabincell{c}{Antenna pitch} & $d = \frac{\lambda}{2}=\frac{3}{560}$ m \\
		\hline
		\tabincell{c}{The noise power at receivers} & $\sigma^{2}=-55$ dBm \\
		\hline
        \tabincell{c}{energy harvesting efficiency} & $\eta=50\%$ \\
		\hline
        \tabincell{c}{Constant scaling coefficient} & $c'=\frac{4}{3}$\\
		\hline
        \tabincell{c}{Number of clusters} & $C = 3$\\
		\hline
        \tabincell{c}{Transmit/Receive antenna  gain at the BS} &  $G_{\text{t}}=G_{\text{r}}=0$ dBm\\
        \hline \tabincell{c}{convergence accuracy}
         & \tabincell{c}{$\epsilon_{1}=10^{-3}$ and  $\epsilon_{2}=\epsilon_{3}=10^{-2}$} \\
		\hline
\end{tabular}
\end{table}

Fig. \ref{Fig.2} depicts the convergence performance of the proposed PTL algorithm. As can be observed, the proposed PTL algorithm can monotonically converge to the stable solutions within 30 iterations, which is expected as at least stationary point solutions are obtained for each subproblem. It is also shown that increasing the number of scatters has little impact on the network performance. This is because, in high-frequency near-field communications, the NLoS channel components follow the cascaded two-hop propagation, which suffers the double path loss and therefore accounts for a minuscule proportion of the whole channel. Moreover, it is observed that increasing $M$ significantly degrades the power consumption at the BS, which is due to the fact that enlarging $M$ introduces more DoFs for the BS, so a more accurate beamfocusing design can be achieved. Besides, it is presented that increasing $M_{\text{MR}}$ slightly reduces transmit power, which validates that the PTL-enabled hybrid beamforming can achieve comparable performance to the fully-digital antenna with a small number of RF chains.


To demonstrate the performance of the proposed algorithms, two baseline schemes are considered in this paper:
\begin{itemize}
  \item \textbf{Fully-digital}: In the fully-digital scheme, each antenna at the BS is equipped with a RF chain. Let us denote the fully-digital beamforming vector as $\mathbf{w}_{\text{FD},k}=\mathbf{P}\mathbf{w}_{k}$, the optimal rank-one fully-digital beamformer $\mathbf{W}_{\text{FD},k}=\mathbf{w}_{\text{FD},k}\mathbf{w}_{\text{FD},k}^{H}$ can be obtained by employing the SDR technique and Proposition \ref{Proposition_3}.
  \item \textbf{Two-stage-ZF}: In the two-stage zero-forcing (two-stage-ZF) scheme, the analog beamformer is determined similar to the low-complexity two-stage algorithm. The baseband information digital beamformers are designed by the ZF criteria, i.e., $\mathbf{W}_{\text{I}}=\mathbf{P}^{H}\mathbf{H}_{\text{I}}(\mathbf{H}_{\text{I}}^{H}
      \mathbf{P}\mathbf{P}^{H}\mathbf{H}_{\text{I}})^{-1}\bm{\Lambda}$, where $\mathbf{H}_{\text{I}}=[\mathbf{h}_{\text{ID},1},\cdots,\mathbf{h}_{\text{ID},K}]$. Then, the power allocation matrix $\bm{\Lambda}$ can be optimally obtained by solving the resultant convex problem.
  \item \textbf{Two-stage-IEA}: In the two-stage information-energy alignment (two-stage-IEA) scheme, the baseband digital beamformers and the former $K+L$ columns of the analog beamformer is designed in the same way as the proposed low-complexity two-stage algorithm. However, the latter $M_{\text{RF}}-K-L$ columns of the analog beamformer is aligned to all the IDs and EHs, i.e., $\mathbf{p}_{j}=\frac{1}{|\mathbf{d}|}\odot\mathbf{d}, (K+L\leq j\leq M_{\text{RF}})$, where $\mathbf{d}=\sum_{k=1}^{K}\mathbf{a}(d_{k},\theta_{k})+\sum_{l=1}^{L}\mathbf{b}(d_{l},\theta_{l})$.
\end{itemize}

\begin{figure}[t]
\centering
\begin{minipage}[t]{0.45\textwidth}
\centering
\includegraphics[width=1\textwidth]{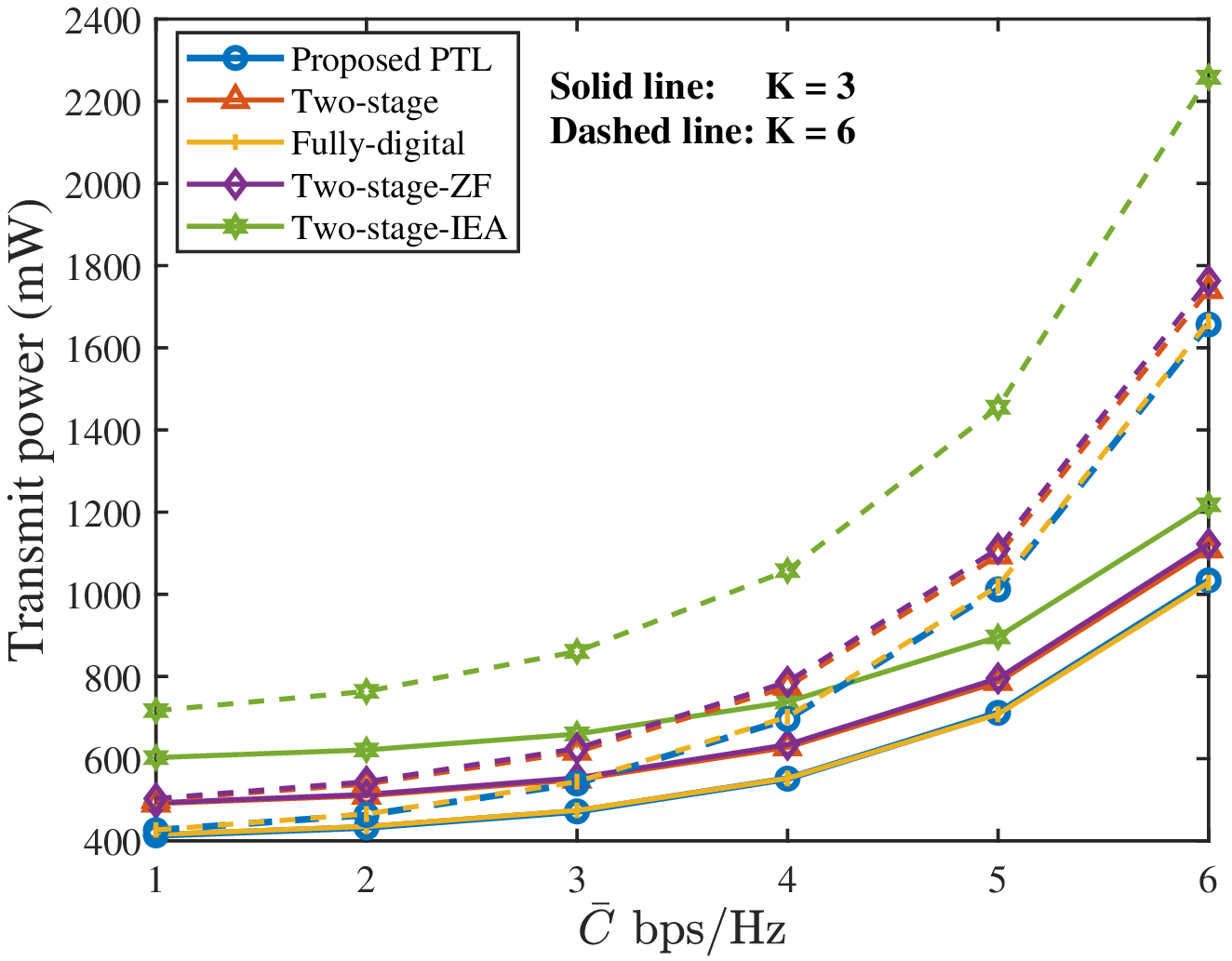}
\caption{Transmit power versus the QoS requirement for different number of IDs with $L = 3$, $\bar{Q}=20$ mW, and $r_{\text{i}}=r_{\text{e}}=20$ m.}
\label{Fig.3}
\end{minipage}\qquad
\begin{minipage}[t]{0.45\textwidth}
\centering
\includegraphics[width=1\textwidth]{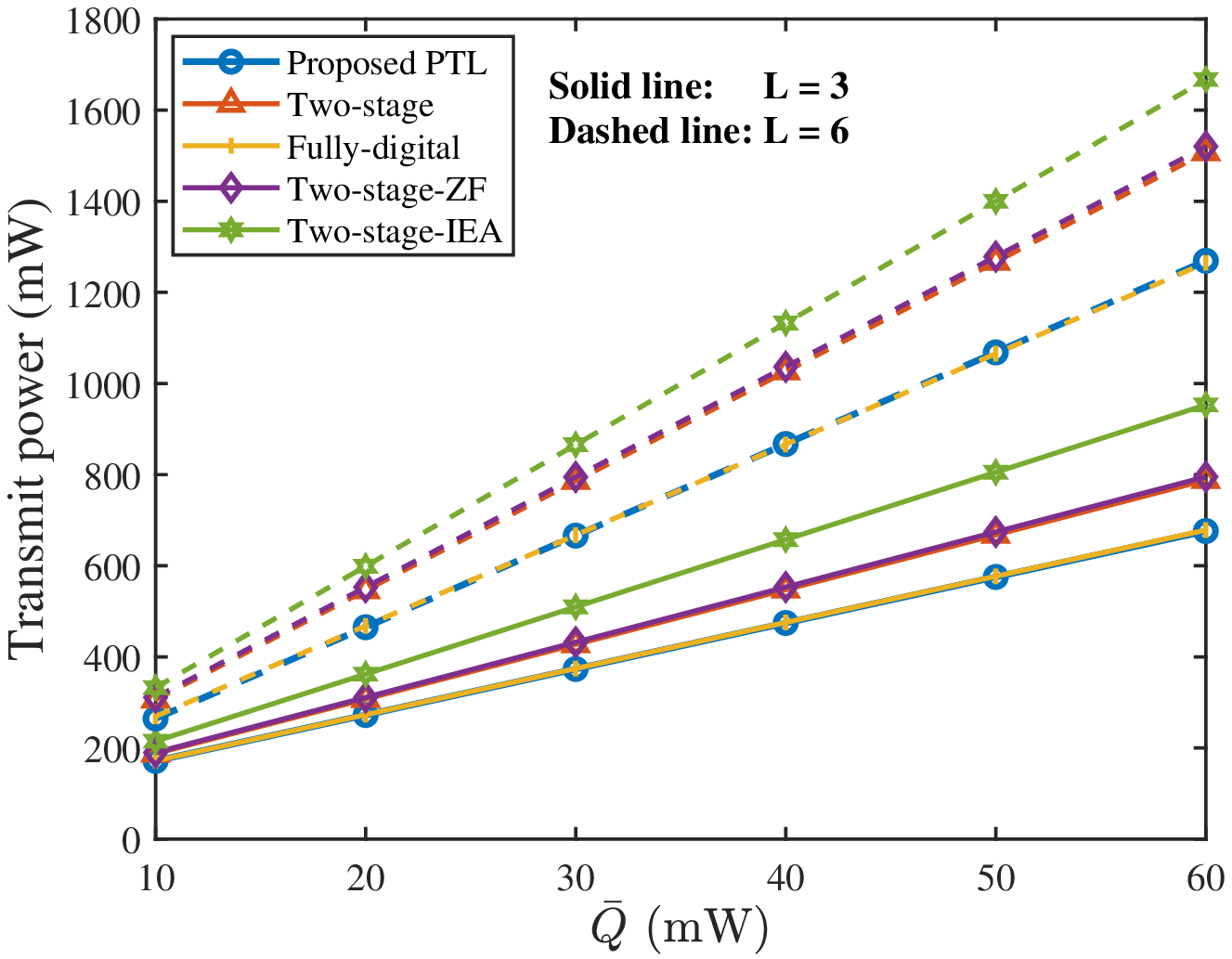}
\caption{Transmit power versus the RF receive power requirement for different number of EHs with $K = 3$, $\bar{C}=1$ bps/Hz, and $r_{\text{i}}=r_{\text{e}}=20$ m.}
\label{Fig.4}
\end{minipage}
\end{figure}
From Fig. \ref{Fig.3} and Fig. \ref{Fig.4}, we can observe that the proposed PTL algorithm can achieves the near-optimal performance compared to the optimal fully-digital scheme, which can be expected because the optimal analog beamformer and the baseband digital beamformers are guarabteed for the corresponding subproblems. It can be also found that the performance achieved by the low-complexity two-stage algorithm is only marginally inferior to the PTL algorithm, which verifies the practicality of the two-stage algorithm as it only possesses the complexity of $\mathcal{O}(M_{\text{RF}}^{3.5})$. Meanwhile, it is presented that the two-stage-ZF scheme realizes a comparable performance to the two-stage algorithm, which indicates that the zero-forcing beams are the near-optimal solutions for the baseband digital beamformers in near-field SWIPT. Furthermore, the two-stage-IEA behaves the worst performance, which is because EHs require significantly higher signal power than that of the signal for IDs, such that the analog beamformer should tend to be aligned to EHs instead of both IDs and EHs. We still observe that increasing the QoS requirement of IDs, RF receive power requirement of EHs, and the number of IDS/EHs will increase the total power consumption at the BS. It is due to the fact that the above changes improve the receive power level at the receivers, which forces the BS to raise the transmit power to satisfy the demands of the network.

\begin{figure}[t]
\centering
\begin{minipage}[t]{0.45\textwidth}
\centering
\includegraphics[width=1\textwidth]{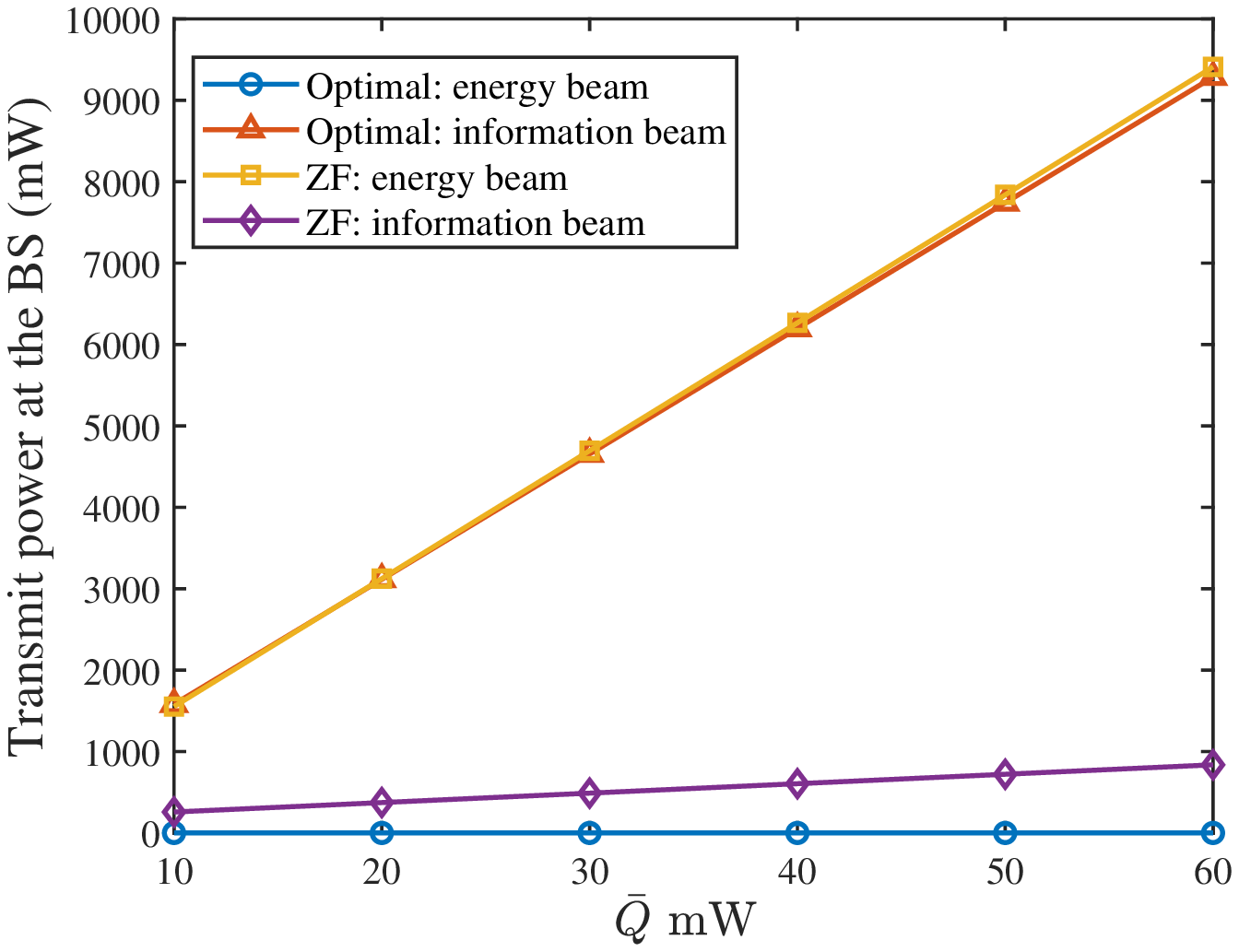}
\caption{Near-field and far-field energy beam comparison with $K=3$, $L=3$, and $\bar{C}=1$ bps/Hz.}
\label{Fig.5}
\end{minipage}\qquad
\begin{minipage}[t]{0.45\textwidth}
\centering
\includegraphics[width=1\textwidth]{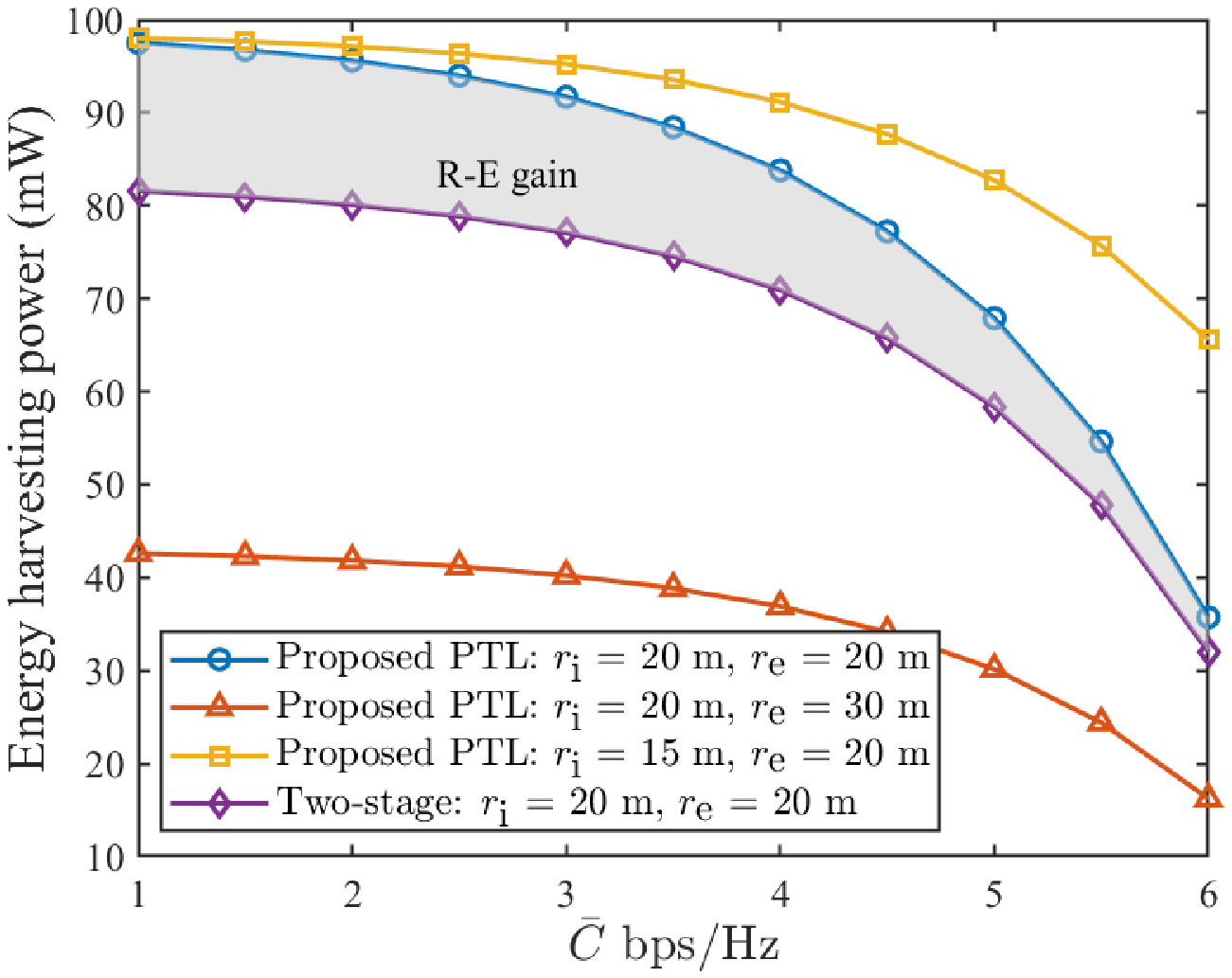}
\caption{R-E region characterization under the 1000 mW total power consumption with $K=3$, $L=3$, and $\bar{Q}=20$ mW.}
\label{Fig.6}
\end{minipage}
\end{figure}

Due to the fact that the analog beamformer is independent of the necessity of the energy beams, we adopt the random analog beamformer in Fig. \ref{Fig.5}, which plots the required information/energy beam power for the different baseband digital beamforming schemes. In particular, the optimal and the ZF baseband digital beamforming can be obtained from the aforementioned two-stage and two-stage-ZF schemes. As can be observed, under the optimal baseband digital beamforming design, no energy beam is required at the BS, which is consistent with the conclusion in Proposition \ref{Proposition_3}. However, when we employ the sub-optimal ZF baseband digital beamforming scheme, A large proportion of the power needs to be  allocated to the energy beams. An intuitive explanation for this interesting phenomenon is that the under the suboptimal ZF scheme, the information beams are mutually constrained, which restricts the flexibility of the information beam design and therefore struggles to fulfill the energy harvesting requirements. As a remedy, more transmit power should be injected into the energy beams, thus satisfying the RF receive power requirements at the EHs.

In Fig. \ref{Fig.6}, we plot the R-E region under a fixed total power consumption, where the PTL algorithm is properly modified to maximize the minimum RF receive power subject to the QoS constraints. It can be observed that the achievable rate monotonically degrades with an increase in harvested energy. This is because if the QoS requirement decreases, more transmit power will be allocated to enhance energy harvesting. It is also shown that the proposed PTL algorithm achieves a larger R-E region than the two-stage algorithm, which is because that the analog beamformer is optimized based on the baseband beamformers in the PTL algorithm, whereas the analog beamformer is determined regardless of baseband beamformers in the two-stage algorithm. We refer to this superiority as the R-E gain of the PTL algorithm over the two-stage algorithm, which is realized by the joint analog and baseband beamforming optimization carried out by the PTL algorithm. Moreover, we can observe that increasing the distance between the IDS/EHs and the BS leads to a diminution of the R-E region, which is because the IDS/EHs suffer severer path loss than before, and more power should be transmitted to compensate for propagation loss.

\begin{figure}[t]
\centering
\begin{minipage}[t]{0.45\textwidth}
\centering
\includegraphics[width=1\textwidth]{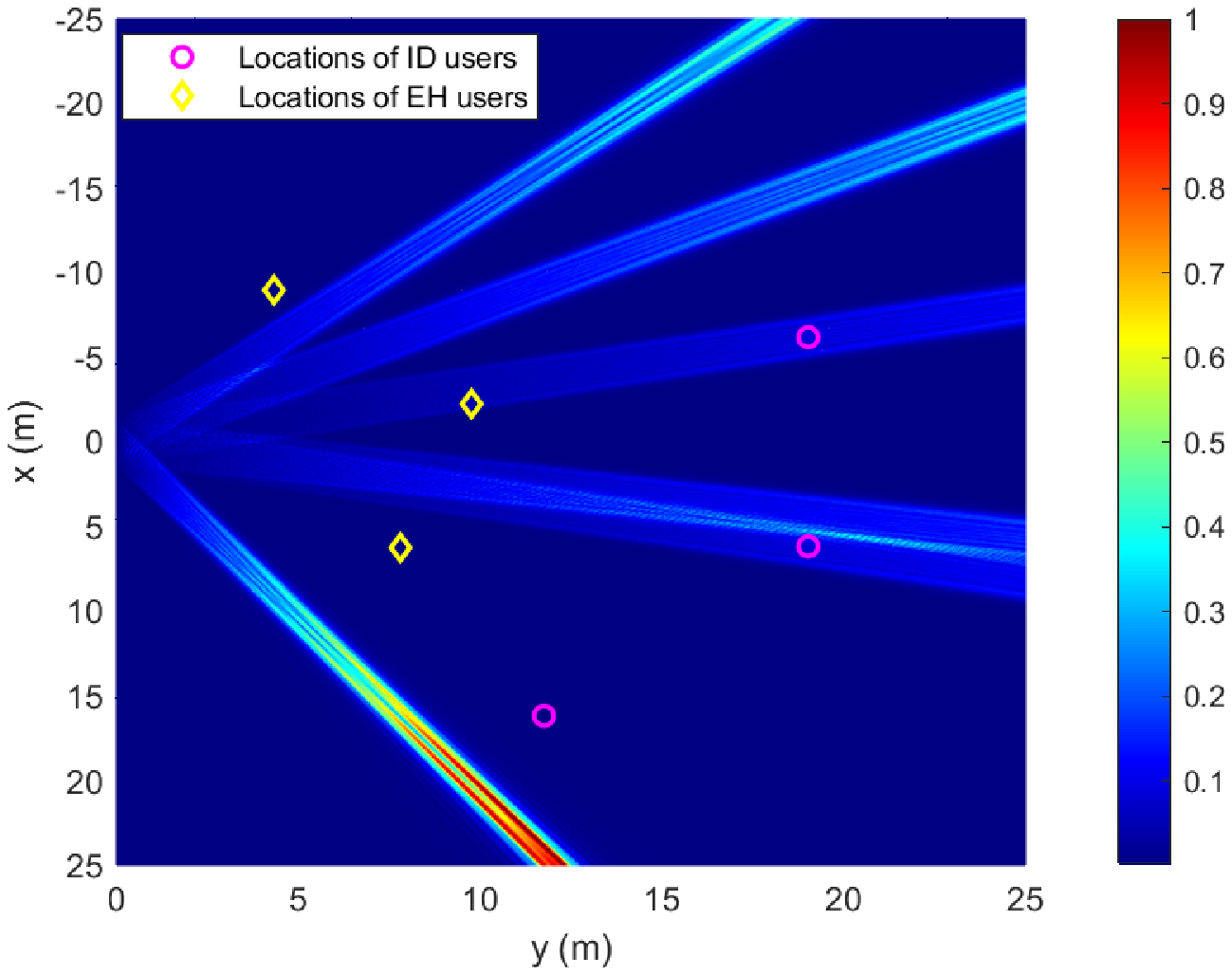}
\caption{Normalized transmit signal power spectrum in far field with $K = 3$, $L = 3$, $\bar{C}=3$ bps/Hz, $\bar{Q}=20$ mW, and $r_{\text{i}}=20$ m, and $r_{\text{e}}=10$ m..}
\label{Fig.7}
\end{minipage}\qquad
\begin{minipage}[t]{0.45\textwidth}
\centering
\includegraphics[width=1\textwidth]{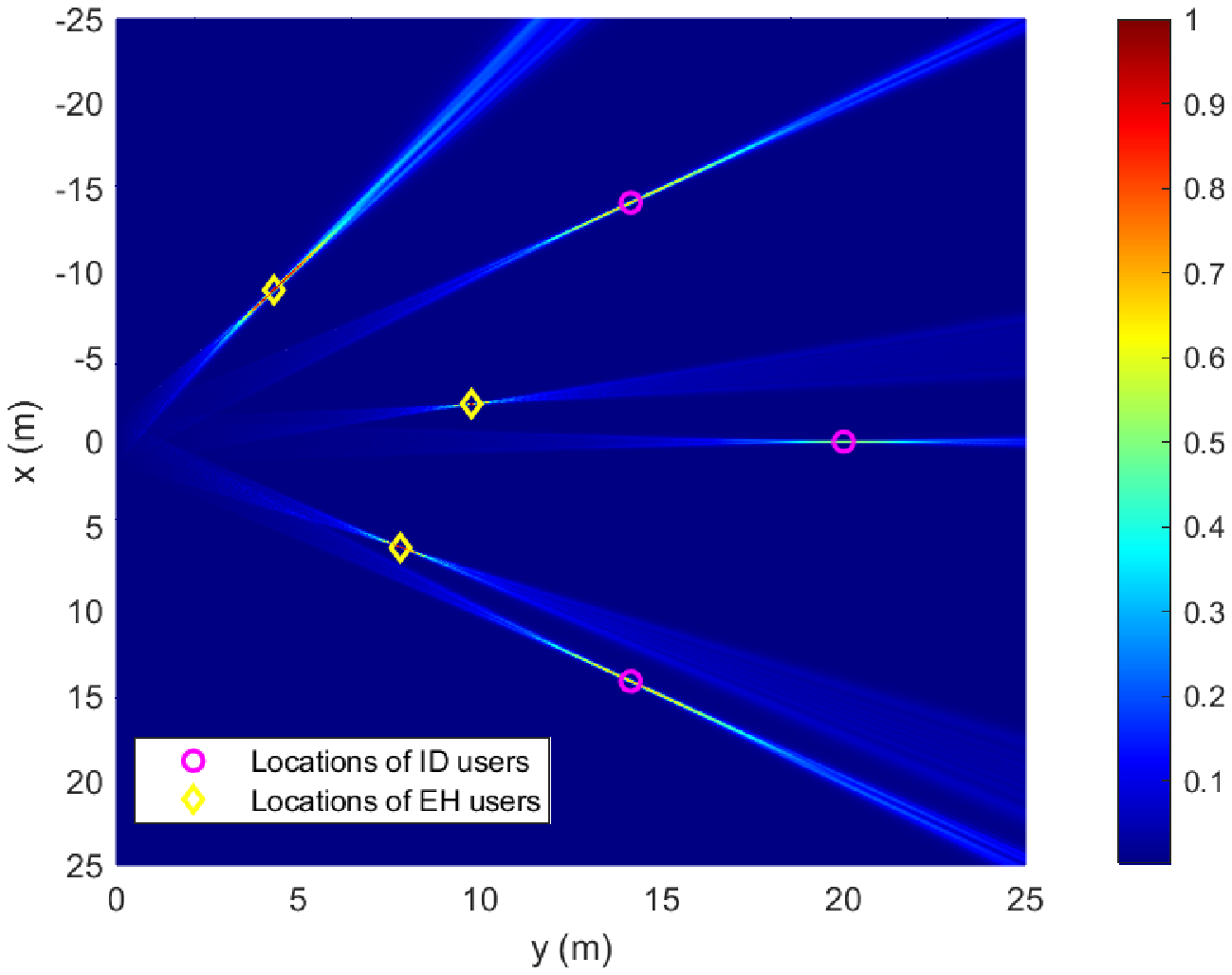}
\caption{Normalized transmit signal power spectrum in near field with $K = 3$, $L = 3$, $\bar{C}=3$ bps/Hz, $\bar{Q}=20$ mW, and $r_{\text{i}}=20$ m, and $r_{\text{e}}=10$ m.}
\label{Fig.8}
\end{minipage}
\end{figure}

In Fig. \ref{Fig.7} and Fig. \ref{Fig.8}, the normalized signal power spectrum achieved by the far-field beam steering and near-field beamfocusing are illustrated. It can be observed that the far-field beam steering cannot achieve signal power enhancement at the locations of IDs/EHs. This is caused by the far-field beam pattern mismatch. Specifically, since all the users are located in the near-field region, the far-field planar wave channel cannot exactly characterize the location information of the receivers, so the far-field beam steering cannot always satisfy the QoS and energy harvest requirements of the near-field SWIPT network. In contrast, Fig. \ref{Fig.8} demonstrates that the spherical wave based near-field beamfocusing can accurately achieve the signal enhancement at the specific points. Moreover, even though only the information beams are applied, it is available to strengthen the signal power at all the users. It unveils an interesting phenomenon that signal power focusing over a large number of free-space points can be achieved using a small number of beamfocusing vectors.

\section{Conclusion}\label{Section_6}
A practical SWIPT framework was investigated for near-field networks, where hybrid beamforming architecture was adopted to achieve the information transmission and energy harvesting. A transmit power minimization problem was formulated via jointly optimizing the analog beamformer, baseband digital information/energy beamformers, and the number of the dedicated energy beams, subject to the QoS requirements and RF receive power constraints. With any given analog beamformer, a generic rank-one solution construction approach was proposed, which guarantees the global optimality of the obtained baseband digital beamformers. By observing the structure of the optimal baseband digital beamformers, it was demonstrated that the number of the dedicated energy beams required for near-field SWIPT is zero. Accordingly, a PTL algorithm was developed to jointly optimize the analog beamformer and baseband digital information beamformer. To decrease the signal processing overhead at the BS, a two-stage algorithm was proposed, which achieved comparable performance to the PTL algorithm with much low computational complexity. The effectiveness of the proposed near-field SWIPT framework was verified by the numerical results. It was also unveiled that the single beamformer was able to achieve the energy focusing on multiple locations, which thus benefiting near-field SWIPT.

\section*{Appendix A: Proof of Proposition \ref{Proposition_1}}
Recall the results in \eqref{15}, we can readily know that all the $\mathbf{\hat{W}}_{k}$ ($k \in \mathcal{K}_{\text{ID}}$) are positive semidefinite and rank-one matrices. Meanwhile, since the problem (P1.2) can be rewritten as
\begin{subequations}
\begin{align}
\tag{A-1a}\label{A-1a} \quad \min\limits_{\mathbf{W}_{k},\mathbf{V}}\quad & \text{Tr}\left(\mathbf{\bar{P}}^{\text{g}}\left(\sum_{k=1}^{K}\mathbf{W}_{k}+\mathbf{V}\right)\right)\\
\tag{A-1b}\label{A-1b} \text{s.t.} \quad & \frac{\text{Tr}(\mathbf{\bar{H}}_{\text{ID},k}\mathbf{W}_{k})}{\bar{\gamma}_{k}}-
\sum_{i\neq k}^{K}\text{Tr}(\mathbf{\bar{H}}_{\text{ID},k}\mathbf{W}_{i})-\text{Tr}(\mathbf{\bar{H}}_{\text{ID},k}\mathbf{V})\geq \sigma^{2}, \ \  k\in \mathcal{K}_{\text{ID}},\\
\tag{A-1c}\label{A-1c}   &\text{Tr}\left(\mathbf{\bar{H}}_{\text{EH},l}\left(\sum_{k=1}^{K}\mathbf{W}_{k}+\mathbf{V}\right)\right) \geq \frac{\bar{Q}_{l}}{\eta},\ \  l\in \mathcal{K}_{\text{EH}},\\
\tag{A-1d}\label{A-1d}   &\mathbf{W}_{k}\succeq\mathbf{0},\ \ \text{rank}(\mathbf{W}_{k}) = 1,\ \   k\in \mathcal{K}_{\text{ID}},\\
\tag{A-1e}\label{A-1e}   & \mathbf{V}\succeq\mathbf{0},
\end{align}
\end{subequations}
it readily knows that the constructed rank-one solutions $\{\mathbf{\hat{W}}_{k},\mathbf{\hat{V}}\}$ guarantee to achieve the same objective value as the original optimal solutions $\{\mathbf{W}_{k}^{*},\mathbf{V}^{*}\}$, while satisfying the constraint \eqref{14c}. Then, we prove $\{\mathbf{\hat{W}}_{k},\mathbf{\hat{V}}^{*}\}$ also satisfy the constraints \eqref{14b} and \eqref{14e}. Based on \eqref{15}, we can obtain following equality
\begin{align}\tag{A-2}\label{A-2}
\mathbf{\bar{h}}_{\text{ID},k}^{H}\mathbf{\hat{W}}_{k}\mathbf{\bar{h}}_{\text{ID},k}=\frac
{\mathbf{\bar{h}}_{\text{ID},k}^{H}\mathbf{W}_{k}^{*}\mathbf{\bar{h}}_{\text{ID},k}
\mathbf{\bar{h}}_{\text{ID},k}^{H}(\mathbf{W}_{k}^{*})^{H}\mathbf{\bar{h}}_{\text{ID},k}}
{\mathbf{\bar{h}}_{\text{ID},k}^{H}\mathbf{W}_{k}^{*}\mathbf{\bar{h}}_{\text{ID},k}}
=
\mathbf{\bar{h}}_{\text{ID},k}^{H}\mathbf{W}_{k}^{*}\mathbf{\bar{h}}_{\text{ID},k}.
\end{align}
Then, it is easily shown
\begin{align}\tag{A-3}\label{A-3}
\text{Tr}(\mathbf{\bar{H}}_{\text{ID},k}\mathbf{\hat{V}})=
\sum_{k=1}^{K}\mathbf{\bar{h}}_{\text{ID},k}^{H}\mathbf{\hat{W}}_{k}\mathbf{\bar{h}}_{\text{ID},k}+
\text{Tr}(\mathbf{\bar{H}}_{\text{ID},k}\mathbf{V}^{*})-
\sum_{k=1}^{K}\mathbf{\bar{h}}_{\text{ID},k}^{H}\mathbf{W}_{k}^{*}\mathbf{\bar{h}}_{\text{ID},k}
=\text{Tr}(\mathbf{\bar{H}}_{\text{ID},k}\mathbf{V}^{*}).
\end{align}
With \eqref{A-2}, \eqref{A-3}, and the fact that $\{\mathbf{W}_{k}^{*},\mathbf{V}^{*}\}$ are feasible to problem (P1.2), we have
\begin{align}\nonumber
&\frac{\text{Tr}(\mathbf{\bar{H}}_{\text{ID},k}\mathbf{\hat{W}}_{k})}{\bar{\gamma}_{k}}
=\frac{\text{Tr}(\mathbf{\bar{H}}_{\text{ID},k}\mathbf{W}_{k}^{*})}{\bar{\gamma}_{k}}\\ \tag{A-4}\label{A-4}
&\quad\geq
\sum_{i\neq k}^{K}\text{Tr}(\mathbf{\bar{H}}_{\text{ID},k}\mathbf{W}_{i}^{*})+
\text{Tr}(\mathbf{\bar{H}}_{\text{ID},k}\mathbf{V}^{*})+ \sigma^{2}=
\sum_{i\neq k}^{K}\text{Tr}(\mathbf{\bar{H}}_{\text{ID},k}\mathbf{\hat{W}}_{i})+
\text{Tr}(\mathbf{\bar{H}}_{\text{ID},k}\mathbf{\hat{V}})+ \sigma^{2}.
\end{align}
Thus, the $\{\mathbf{\hat{W}}_{k},\mathbf{\hat{V}}\}$ meet the QoS constraint \eqref{14b}. On the other hand, for any complex vector $\mathbf{x}\in\mathbb{C}^{M_{\text{RF}}\times1}$, we can obtain
\begin{align}\tag{A-5}\label{A-5}
\mathbf{x}^{H}(\mathbf{W}_{k}^{*}-\mathbf{\hat{W}}_{k})\mathbf{x}=
\mathbf{x}^{H}(\mathbf{W}_{k}^{*})\mathbf{x}-\frac
{\mathbf{x}^{H}\mathbf{W}_{k}^{*}\mathbf{\bar{h}}_{\text{ID},k}
\mathbf{\bar{h}}_{\text{ID},k}^{H}(\mathbf{W}_{k}^{*})^{H}\mathbf{x}}
{\mathbf{\bar{h}}_{\text{ID},k}^{H}\mathbf{W}_{k}^{*}\mathbf{x}}.
\end{align}
By checking the Cauchy-Schwarz inequality, it holds that
\begin{align}\tag{A-6}\label{A-6}
(\mathbf{x}^{H}\mathbf{W}_{k}^{*}\mathbf{x})(\mathbf{\bar{h}}_{\text{ID},k}^{H}\mathbf{W}_{k}^{*}\mathbf{\bar{h}}_{\text{ID},k})\geq
|\mathbf{x}^{H}\mathbf{W}_{k}^{*}\mathbf{\bar{h}}_{\text{ID},k}|^{2}.
\end{align}
Substituting \eqref{A-6} into \eqref{A-5}, we have
\begin{align}\tag{A-7}\label{A-7}
\mathbf{x}^{H}(\mathbf{W}_{k}^{*}-\mathbf{\hat{W}}_{k})\mathbf{x}\geq0,
\end{align}
which implies that $\mathbf{W}_{k}^{*}-\mathbf{\hat{W}}_{k}$ is a positive semidefinite matrix. Moreover, since $\mathbf{V}^{*}$ is feasible to problem (P1.2), $\mathbf{V}^{*}\succeq\mathbf{0}$ always holds. Thus, the constructed $\mathbf{\hat{V}}=\mathbf{W}_{k}^{*}-\mathbf{\hat{W}}_{k}+\mathbf{V}^{*}$ is also positive semidefinite. This completes the proof.

\section*{Appendix B: Proof of Proposition \ref{Proposition_2}}
Since $\{\mathbf{\hat{W}}_{k},\mathbf{\hat{V}}\}$ is feasible to the problem (P1.2), it always holds that
\begin{align}\nonumber
&\frac{\text{Tr}(\mathbf{\bar{H}}_{\text{ID},k}\mathbf{\tilde{W}}_{k})}{\bar{\gamma}_{k}}
=\frac{\text{Tr}(\mathbf{\bar{H}}_{\text{ID},k}(\mathbf{\hat{W}}_{k}+\alpha_{k}\mathbf{\hat{V}}))}{\bar{\gamma}_{k}}
\geq\frac{\text{Tr}(\mathbf{\bar{H}}_{\text{ID},k}\mathbf{\hat{W}}_{k})}{\bar{\gamma}_{k}}
\\ \tag{B-1}\label{B-1}
&\qquad\overset{a}{\geq}
\sum_{i\neq k}^{K}\text{Tr}(\mathbf{\bar{H}}_{\text{ID},k}\mathbf{\hat{W}}_{i})+
\text{Tr}(\mathbf{\bar{H}}_{\text{ID},k}\mathbf{\hat{V}})+ \sigma^{2}=
\sum_{i\neq k}^{K}\text{Tr}(\mathbf{\bar{H}}_{\text{ID},k}\mathbf{\tilde{W}}_{k})+ \sigma^{2},
\end{align}
where inequality $a$ holds because $\sum_{k=1}^{K}\mathbf{\tilde{W}}_{k} = \sum_{k=1}^{K}(\mathbf{\hat{W}}_{k}+\alpha_{k}\mathbf{\hat{V}})=\sum_{k=1}^{K}\mathbf{\hat{W}}_{k}+\mathbf{\hat{V}}$. Thus, the constructed solutions $\{\mathbf{\tilde{W}}_{k}\}$ satisfy \eqref{14b}. Meanwhile, it is easily known that solutions $\{\mathbf{\tilde{W}}_{k}\}$ are positive semidefinite, and achieve the same performance as $\{\mathbf{\hat{W}}_{k},\mathbf{\hat{V}}\}$ while ensuring that \eqref{14c} is satisfied. As a result, $\{\mathbf{\tilde{W}}_{k}\}$ are feasible to the rank-one relaxed problem. Then, we prove the optimality of $\{\mathbf{\tilde{W}}_{k}\}$ via the proof by contradiction. Specifically, we assume there exist the optimal solutions $\{\mathbf{\tilde{W}}_{k}^{\text{o}}\}$ different from $\{\mathbf{\tilde{W}}_{k}\}$. Thus, it readily knows that
\begin{align}\tag{B-2}\label{B-2}
\mathfrak{O}_{\text{P}1.2}^{*}(\mathbf{\tilde{W}}_{k}^{\text{o}})<
\mathfrak{O}_{\text{P}1.2}^{*}(\mathbf{\tilde{W}}_{k})=
\mathfrak{O}_{\text{P}1.2}^{*}(\mathbf{\hat{W}}_{k},\mathbf{\hat{V}}),
\end{align}
where $\mathfrak{O}_{\text{P}1.2}^{*}(\mathbf{X})$ denotes the optimal objective value of the problem (P1.2) achieved by the solution $\mathbf{X}$. Meanwhile, we can always construct a corresponding solutions $\{\mathbf{\tilde{W}}_{k}^{\text{o}},\mathbf{0}\}$ that are feasible to the problem (P1.2), i.e.,
\begin{align}\tag{B-3}\label{B-3}
\mathfrak{O}_{\text{P}1.2}^{*}(\mathbf{\tilde{W}}_{k}^{\text{o}},\mathbf{0})=
\mathfrak{O}_{\text{P}1.2}^{*}(\mathbf{\tilde{W}}_{k}^{\text{o}}).
\end{align}
With results of \eqref{B-2} and \eqref{B-3}, we can obtain $\mathfrak{O}_{\text{P}1.2}^{*}(\mathbf{\tilde{W}}_{k}^{\text{o}},\mathbf{0})<
\mathfrak{O}_{\text{P}1.2}^{*}(\mathbf{\hat{W}}_{k},\mathbf{\hat{V}})$. However, this contradicts the fact that $\{\mathbf{\hat{W}}_{k},\mathbf{\hat{V}}\}$ are optimal solutions to the problem (P1.2). Thus, $\{\mathbf{\tilde{W}}_{k}^{\text{o}}\}$ do not exist and the constructed solutions $\{\mathbf{\tilde{W}}_{k}\}$ are optimal to the rank-one relaxed problem.


\begin{thebibliography}{99}

\bibitem{F.Boccardi_5G_magazine}
F. Boccardi, R. W. Heath, A. Lozano, T. L. Marzetta, and P. Popovski, ``Five disruptive technology directions for 5G,''\textit{IEEE Commun. Mag.}, vol. 52, no. 2, pp. 74--80, Feb. 2014.

\bibitem{6G_samsung}
Samsung Research, ``6G: The next hyper connected experience for all,'' Samsung, While Paper, 2020. [Online]. Available:
https://research.samsung.com/next-generation-communications

\bibitem{Y.Liu_RIS}
Y. Liu et al., ``Reconfigurable intelligent surfaces: Principles and opportunities,'' \textit{IEEE Commun. Surveys Tuts.}, vol. 23, no. 3, pp. 1546--1577, Third quarter 2021.

\bibitem{I.Krikidis_SWIPT}
I. Krikidis, S. Timotheou, S. Nikolaou, G. Zheng, D. W. K. Ng, and R. Schober, ``Simultaneous wireless information and power transfer in modern communication systems,'' \textit{IEEE Commun. Mag.}, vol. 52, no. 11, pp. 104--110, Nov. 2014.



\bibitem{S.Bi_WPT_mag}
S. Bi, C. K. Ho, and R. Zhang, ``Wireless powered communication: Opportunities and challenges,'' \textit{IEEE Wireless Commun.}, vol. 53, no. 4, pp. 117--125.

\bibitem{L.Liu_SWIPT}
L. Liu, R. Zhang, and K. -C. Chua, ``Wireless information transfer with opportunistic energy harvesting,'' \textit{IEEE Trans. Wireless Commun.}, vol. 12, no. 1, pp. 288--300, Jan. 2013.


\bibitem{Y.Liu_NF_mag}
Y. Liu, J. Xu, Z. Wang, X. Mu, and L. Hanzo, ``Near-field communications: What will be different?'' [Online]. Available: https://arxiv.org/abs/2303.04003


\bibitem{J.Xu_NF}
J. Xu, X. Mu, and Y. Liu, ``Exploiting STAR-RISs in near-field communications,'' [Online]. Available: https://arxiv.org/abs/2211.15777

\bibitem{C.Huang_NF_channel}
L. Wei, C. Huang, et al., ``Tri-polarized holographic MIMO surfaces for near-field communications: Channel modeling and precoding design,'' \textit{IEEE Trans. Wireless Commun.}, early access, doi: 10.1109/TWC.2023.3266298.

\bibitem{R.Ji_DoF}
R. Ji, S. Chen, C.Huang, et al., ``Extra DoF of near-field holographic MIMO communications leveraging evanescent waves,'' \textit{IEEE Wireless Commun. Lett.}, vol. 12, no. 4, pp. 580--584, Apr. 2023.

\bibitem{H.Zhang_NF_mag}
H. Zhang, N. Shlezinger, F. Guidi, D. Dardari, and Y. C. Eldar, ``6G wireless communications: From far-field beam steering to near-field beam focusing,'' \textit{IEEE Commun. Mag.}, vol. 61, no. 4, pp. 72--77, Apr. 2023.

\bibitem{H.Zhang_NF_mag2}
H. Zhang, N. Shlezinger, F. Guidi, D. Dardari, M. F. Imani, and Y. C. Eldar, ``Near-field wireless power transfer for 6G internet of everything mobile networks: Opportunities and challenges,''  \textit{IEEE Commun. Mag.}, vol. 60, no. 3, pp. 12--18, Mar. 2022.

\bibitem{M.Cui_NF_mag}
M. Cui, Z. Wu, Y. Lu, X. Wei, and L. Dai, ``Near-field MIMO communications for 6G: Fundamentals, challenges, potentials, and future Directions,'' \textit{IEEE Commun. Mag.}, vol. 61, no. 1, pp. 40--46, Jan. 2023.

\bibitem{C.Huang_NF1}
X. Gan, C. Huang, Z. Yang, C. Zhong, and Z. Zhang, ``Near-field localization for holographic RIS assisted mmWave systems,'' \textit{IEEE Commun. Lett.}, vol. 27, no. 1, pp. 140--144, Jan. 2023.


\bibitem{L.R.Varshney_SWIPT}
L. R. Varshney, ``Transporting information and energy simultaneously,'' in \textit{Proc. IEEE Int. Symp. Inf. Theory},  pp. 1612--1616, Jul. 2008.


\bibitem{R.Zhang_SWIPT}
R. Zhang and C. K. Ho, ``MIMO Broadcasting for simultaneous wireless information and power transfer,'' \textit{IEEE Trans. Wireless Commun.}, vol. 12, no. 5, pp. 1989--2001, May. 2013.


\bibitem{J.Xu_energy_beam1}
J. Xu, L. Liu, and R. Zhang, ``Multiuser MISO beamforming for simultaneous wireless information and power transfer,'' \textit{IEEE Trans. Signal Process.}, vol. 62, no. 18, pp. 4798--4810, Sep. 2014.



\bibitem{Q.Wu_SWIPT_wcl}
Q. Wu and R. Zhang, ``Weighted sum power maximization for intelligent reflecting surface aided SWIPT,'' \textit{IEEE Wireless Commun. Lett.}, vol. 9, no. 5, pp. 586--590, May. 2020.

\bibitem{Y.Liu_SWIPT_NOMA}
Y. Liu, Z. Ding, M. Elkashlan, and H. V. Poor, ``Cooperative non-orthogonal multiple access with simultaneous wireless information and power transfer,'' \textit{IEEE J. Sel. Areas Commun.}, vol. 34, no. 4, pp. 938--953, Apr. 2016.

\bibitem{H.Zhang_NF}
H. Zhang, N. Shlezinger, et al, ``Beam focusing for near-field multiuser MIMO communications,'' \textit{IEEE Trans. Wireless Commun.}, vol. 21, no. 9, pp. 7476--7490, Sep. 2022.

\bibitem{LDMA}
Z. Wu and L. Dai, ``Multiple access for near-field communications: SDMA or LDMA?'' [Online]. Available: https://arxiv.org/abs/2208.06349



\bibitem{NF_analysis}
K. Zhi, C. Pan, H. Ren, K. Chai, C. Wang, R. Schober, and, X. You, ``Performance analysis and low-complexity design for XL-MIMO with near-field spatial non-stationarities,'' [Online]. Available: https://arxiv.org/abs/2304.00172

\bibitem{H.Zhang_channel_estimation}
X. Zhang, Z. Wang, H. Zhang, and L. Yang, ``Near-field channel estimation for extremely large-scale array communications: A model-based deep learning approach,'' \textit{IEEE Commun. Lett.}, vol. 27, no. 4, pp. 1155--1159, Apr. 2023.

\bibitem{C.You_NF_training}
Y. Zhang, X. Wu, and C. You, ``Fast near-field beam training for extremely large-scale array,'' [Online]. Available: https://arxiv.org/abs/2209.14798

\bibitem{C.Wu_NF_training}
C. Wu, C. You, et. al. ``Two-stage hierarchical beam training for near-field communications,'' [Online]. Available: https://arxiv.org/abs/2302.12511

%


%
%

\bibitem{H.Zhang_WPT}
H. Zhang, N. Shlezinger, F. Guidi, D. Dardari, M. F. Imani, and Y. C. Eldar, ``Near-field wireless power transfer with dynamic metasurface antennas,'' in \textit{Proc. 2022 IEEE 23rd Int. Workshop Signal Process. Advances Wireless Commun.}, Oulu, Finland, 2022, pp. 1--5.

\bibitem{E.Demarchou_energy_focusing}
E. Demarchou, C. Psomas, and I. Krikidis, ``Energy focusing for wireless power transfer in the near-field region,'' in \textit{Proc. IEEE Global Commun. Conf.}, Brazil, Dec. 2022, pp. 4106--4110.

\bibitem{H.Zhang_NF_ISWPT}
Q. Yang, H. Zhang, and B. Wang, ``Beamforming design for integrated sensing and wireless power transfer systems,'' \textit{IEEE Commun. Lett.}, vol. 27, no. 2, pp. 600--604, Feb. 2023.





\bibitem{X.Yu_MIMO_Hybrid}
X. Yu, J. -C. Shen, J. Zhang, and K. B. Letaief, ``Alternating minimization algorithms for hybrid precoding in millimeter wave MIMO systems,'' \textit{IEEE J. Sel. Top. Signal Process.}, vol. 10, no. 3, pp. 485--500, Apr. 2016.








\bibitem{Z.Luo_complexity}
Z. Luo, W. Ma, A. M. So, Y. Ye, and S. Zhang, ``Semidefinite relaxation of quadratic optimization problems,''
\textit{IEEE Signal Process. Mag.}, vol. 27, no. 3, pp. 20--34, May. 2010.




\bibitem{strong_duality}
A. Nemirovski, ``Lectures on modern convex optimization,'' Class notes, Georgia Inst. of Technology, Atlanta, 2005.


\vspace{-0.5mm}
\bibitem{Y.Huang_SSDP}
Y. Huang and D. Palomar, ``Rank-constrained separable semidefinite programming with applications to optimal beamforming,'' \textit{IEEE Trans. Signal Process.}, vol. 58, no. 2, pp. 664--678, Feb. 2010.

\vspace{-0.5mm}
\bibitem{PDD}
Q. Shi and M. Hong, ``Penalty dual decomposition method for nonsmooth nonconvex optimization part I: Algorithms and convergence analysis,'' \textit{IEEE Trans. Signal Process.}, vol. 68, pp. 4108--4122, Jun. 2020.

\vspace{-0.5mm}

\bibitem{S.Boyd}
S. Boyd and L. Vandenberghe, \textit{Convex Optimization}. Cambridge, U.K.:
Cambridge Univ. Press, 2004.






%
%
%
%
%
%
%
%
%
%

%

%
%
%
%
%
%
%

%
%

\end{thebibliography}
\end{document}